%% file: _fp076-gonzalez.tex
\documentclass{www2013-accepted}
\usepackage{amsmath}
\usepackage{amsfonts}
\usepackage{array}
\usepackage{amssymb}
\usepackage{graphicx}
\usepackage[hyphens]{url}
\usepackage{multicol}
\usepackage{stfloats}
\usepackage{subfigure}
\usepackage[usenames]{color}

\usepackage{epstopdf}

\usepackage{booktabs}
\usepackage{changes}
\definechangesauthor{Ru}{red}
\definechangesauthor{An}{orange}
\definechangesauthor{Re}{blue}

\newcommand{\eg}{{\it e.g.}}
\newcommand{\ie}{{\it i.e.}}

\newcommand{\bi}{\begin{itemize}}
\newcommand{\ei}{\end{itemize}}
\newcommand {\beq}{\begin{equation}}
\newcommand {\eeq}{\end{equation}}
\newcommand {\be}{\begin{enumerate}}
\newcommand {\ee}{\end{enumerate}}

\begin{document}

\title{Google+ or Google-?\\Dissecting the Evolution of the \emph{\huge New} OSN in its First Year}

\numberofauthors{3} 
\author{
Roberto Gonzalez, Ruben Cuevas \\
\affaddr{Universidad Carlos III de Madrid} \\ \affaddr{ Leganes, Madrid, Spain} \\
\email{\{rgonza1,rcuevas\}@it.uc3m.es} \\
\and
Reza Motamedi, Reza Rejaie \\
\affaddr{University of Oregon} \\ \affaddr{ Eugene, OR, US} \\
\email{\{motamedi,reza\}@cs.uoregon.edu}\\
\and
Angel Cuevas \\
\affaddr{Telecom Sud Paris} \\ \affaddr{ Evry, \^Ile-de-France, France} \\
\email{angel.cuevas\_rumin@it\-sudparis.eu}\\
}


\maketitle

\begin{abstract}

In the era when Facebook and Twitter dominate the market
for social media, Google has introduced Google+ (G+) and
reported a significant growth in its size while others called it
a ghost town. This begs the question that "whether G+ can
really attract a significant number of connected and active
users despite the dominance of Facebook and Twitter?".

This paper tackles the above question by presenting a detailed 
characterization of G+ based on large scale measurements. We identify the 
main components of G+ structure, characterize the key features of their
users and their evolution over time. We then conduct detailed
analysis on the evolution of connectivity and activity
among users in the largest connected component (LCC) of
G+ structure, and compare their characteristics with other
major OSNs. We show that despite the dramatic growth in
the size of G+, the relative size of LCC has been decreasing
and its connectivity has become less clustered. While the
aggregate user activity has gradually increased, only a very
small fraction of users exhibit any type of activity. To our
knowledge, our study offers the most comprehensive characterization
of G+ based on the largest collected data sets.

\end{abstract}



\input{intro-rr2}

\input{overview-rr}

\input{measurements-rr2}

\input{elements-rr2}

\input{activity-rr2}

\input{connectivity-rr2}
\input{related_work}
\input{conclusions}

\input{acknowledgments}

{\small
\bibliographystyle{plain}
\bibliography{G+}
}

\end{document}

%% file: intro-rr2.tex

\section{Introduction}

A significant majority of today's Internet users rely on Facebook and Twitter for their online social interactions. In June of 2011, Google launched a new Online Social Network (OSN), called {\em Google+} (or {\em G+} for short) in order to claim a fraction of the social media market and its associated profit. G+ offers a combination of Facebook- and Twitter-like services in order to attract users from both rivals. There has been several official reports about the rapid growth of G+'s user population (400M users in Sep 2012) \cite{G+_official_stats} while some observers and users dismissed these claims and called G+ a ``ghost town'' \cite{Wall_Street_Journal}. This raises the following important question: {\em ``Can a new OSN such as G+ really attract a significant number of engaged users and be a relevant player in the social media market?"}. A major Internet company such as Google, with many popular services, is perfectly positioned to implicitly or explicitly require (or motivate) its current users to join its OSN. Then, it is interesting to assess to what extent and how Google might have leveraged its position to make users join G+. Nevertheless, any growth in the number of users in an OSN is really meaningful only if the new users adequately connect to the rest of the network (\ie,~become connected) and become active by using some of the offered services by the OSN on a regular basis. 
We also note that today's Internet users are much more savvy about using OSN services and connecting to other users than users a decade ago when Facebook and Twitter became popular. This raises other related questions: {\em ``how has the connectivity and activity of G+ users evolved over time as users have become significantly more experienced about using OSNs?'' and ``whether these evolution patterns exhibit different characteristics compared to earlier major OSNs?''}. These evolution
patterns could also offer an insight on whether users willingly join G+ or are added to the system by Google.


In this paper, we present a comprehensive measurement-based characterization of connectivity and activity among G+ users and their evolution over time in order to shed an insightful light on all the above questions. 
We start by providing a brief overview of G+ in Section \ref{sec:overview}. One of our contributions is our measurement methodology to efficiently capture complete snapshots of G+'s largest connected component (LCC), several large sets of randomly selected users, and all the publicly-visible activities (\ie,~user posts) of LCC users with their associated reactions from other users. To our knowledge, this is one of the largest and more diverse collection of datasets used to characterize an OSN. We describe our datasets in Section \ref{sec:measurements} along with our measurement methodology.

In Section \ref{sec:elements}, using our LLC snapshots, we characterize the evolution of the LCC size during the past year. Furthermore, we leverage the randomly selected users to characterize the relative size of the main components (\ie,~LCC, small partitions, and singletons) of G+ network and the evolution of their relative size over time along with the fraction of publicly visible posts and profile attributes for users in each component. Our results show that while the size of the LCC has increased at an impressive pace over the past year, its relative size has consistently decreased such that the LCC currently makes up only 1/3rd of the network and the rest of the users are mostly singletons. 
The large and growing fraction of singletons appears to be caused by Google's integrated registration process that implicitly creates a G+ account for any new Google account regardless of the user's interest. Furthermore, we discover that LCC users generate most of the public posts and provide a larger number of attributes in their profile. Since LCC users form the most important component of G+ network, we focus the rest of our analysis on the LCC.

We then turn our attention to the publicly visible activity of LCC users and its evolution during the entire lifetime of G+ in Section \ref{sec:act}. We discover that the aggregate number of posts by LCC users and their reactions (namely comments, plusones or reshares) from other users have been steadily increasing over time.
Furthermore, a very small fraction of LCC users generate posts and the posts from an even smaller fraction of these users receive most of the reactions from other users, \ie,~user actions and reactions are concentrated around a very small fraction of LCC users.  The fraction of active LCC users has increased 60 times slower than LCC population.
The comparison of users activity among G+, Twitter and Facebook reveals that G+ users are significantly less active than in the other two OSNs.

Finally, we explore the evolution of connectivity features of the LCC in Section \ref{sec:connect} and show that many of its features have initially evolved but have stabilized in recent months despite the continued significant growth in its population, \ie,~the connectivity features appear to have reached a level of maturity. Interestingly, many connectivity features of the current G+ network have a striking similarity with the same feature in Twitter but are very different from Facebook. More specifically, the fraction of reciprocated edges among LCC users are small (and mostly associated with low degree and non-active users) and the LCC network has become increasingly less clustered. Furthermore, we observe a strong positive correlation between the popularity (\ie,~number of followers) and rate of posts of individual G+ users.
In summary, the similarity of connectivity features for G+ and Twitter networks coupled with  the concentration of posting
activity (and their reaction) on a small fraction of popular users and the small fraction of bidirectional relationships in the LCC indicate that 
G+ network is primarily used for broadcasting information.
%

%% file: overview-rr.tex
\section{Google+ Overview}
\label{sec:overview}

After a few unsuccessful attempts (Buzz \cite{buzz_closure}, Wave \cite{wave_shutdown} and Orkut \cite{official_orkut,orkut_statistics}), Google launched G+ on June 28$^{th}$ 2011 with the intention of becoming a major player in the OSNs market. Users were initially allowed to join by invitation.  On September 20$^{th}$,  G+ became open to public and the G+ Pages service was launched on November 7$^{th}$ 2011 \cite{gpages_website,gpages_announcement}. This service imitates the \emph{Facebook Pages} enabling businesses to connect with interested users. Furthermore, also in November 2011, the registration process was integrated with other Google services (\eg,~Gmail, YouTube) \cite{google_registration2,google_registration}.

G+ features have some similarity to Facebook and Twitter.  Similar to Twitter (and different from Facebook) the relationships in G+ are unidirectional. 
More specifically, user $A$ can follow user $B$ in G+ and view all of $B$'s public posts without requiring  the relationship to be reciprocated. We refer to $A$ as $B$'s \emph{follower} and to $B$ as $A$'s \emph{friend}. Moreover, a user can also control the visibility of a post to a specific subset of its followers by grouping them into {\em circles}. This feature imitates Facebook approach to control visibility of shared content. It is worth noting that this circle-based privacy setting is rather complex for average users to manage and thus unskilled users may not use it properly\footnote{A clear example of this complexity is the diagram provided to guide users to determine their privacy setting in \cite{g+_flowchart}.}.

Each user has a stream (similar to Facebook wall) where any activity performed by the user appears. The main activity of a user is to make a ``post''. A post consists of some (or no) text that may have one or more attached files, called ``attachments''. Each attachment could be a video, a photo or any other file.
Other users can react to a particular post in three different ways: {\em (i) Plusone}:  this is similar to the ``like'' feature in Facebook with which other users can indicate their interest in a post, {\em (ii) Comment}: other users can make comments on a post, and {\em (iii) Reshare}: this feature is similar to a ``retweet'' in Twitter and allows other users resend a post to their followers.

G+ assigns a numerical user ID and a profile to each user. The user ID is a 21-digit integer where the highest order digit is always 1 (\eg,~113104553286769158393). Our examination of the assigned IDs did not reveal any clear strategy for ID assignment (\eg,~based on time or mod of certain numbers). Note that this extremely large ID space (10$^{20}$) is sparsely populated (large distance between user IDs) which in turn makes identifying valid user IDs by generating random numbers impractical. 
Similar to other OSNs, G+ users have a profile that has 17 fields where they can provide a range of information and pointers (\eg,~to their other pages) about themselves. However, providing this information (except for sex) is not mandatory for creating an account and thus users may leave some (or all) attributes in their profile empty. Furthermore, users can limit the visibility of specific attributes by defining them as ``private'' and thus visible to a specific group\footnote{Note that it is not possible to distinguish whether a non visible attribute is private or not specified by the user.}.
For a more detailed description of G+ functionality we refer the reader to \cite{g+_learnmore,g+_support}.



%% file: measurements-rr2.tex
\section{Measurement Methodology and Datasets}
\label{sec:measurements}
This section presents our techniques for data collection (and validation) and then a summary of our datasets that we use for our analysis.

\noindent \textbf{Capturing LCC Structure:}
To capture the connectivity structure of the Largest Connected Component (LCC), we use a few high-degree users as starting seeds and crawl the structure using a breadth-first search (BFS) strategy. Our initial examination revealed that the allocated users IDs are very evenly distributed across the ID space. We leverage this feature to speed up our crawler as follows: We divide the ID space into 21 equal-size zones and assign a crawler to only crawl users whose ID falls in a particular zone. Given user $u$ in zone $i$, the assigned crawler to zone $i$ collects the profile along with the list of friends and followers for user $u$. Any newly discovered users whose ID is in zone $i$ are placed in a queue to be crawled whereas discovered users from other zones are periodically reported to a central coordinator. The coordinator maps all the reported users by all 21 crawlers to their zone and periodically (once per hour) sends a list of discovered users in each zone to the corresponding crawler. This strategy requires infrequent and efficient coordination with crawlers and enables them to crawl their zones in parallel. The crawl of each zone is completed when there is no more users in that zone to crawl. After some tuning, the average rate of discovery for each crawler reached 800K users per hour or 16.8M users per day for the whole system\footnote{LCC-Apr snapshot was collected before this tuning and therefore it took longer.}. With this rate, it takes 4-6 days to capture a full snapshot of the LCC connectivity and users' profiles. Table \ref{tab:lcc-dataset} summarizes the main characteristics of our LCC datasets. We obtained the LCC-Dec snapshot from an earlier study on G+\cite{g+_imc_cha}. We examined the connectivity of all the captured LCC snapshots and verified that all of them form a single connected component.
\begin{table}
\small
\centering
\begin{tabular}{|c||c|c|c|c|c|}\hline
Name & \#nodes &  \#edges & Start Date & Duration (days)\\ \hline\hline
LCC-Dec* & 35.1M &  575M & 11-Nov-12 & 46 \\ \hline
LCC-Apr & 51.8M & 1.1B & 15-Mar-12 &  29\\ \hline
LCC-Aug & 79.2M & 1.6B & 20-Aug-12 & 4\\ \hline
LCC-Sep & 85.3M & 1.7B & 17-Sep-12 & 5\\ \hline
LCC-Oct & 89.8M &  1.8B & 15-Oct-12 & 5\\ \hline
LCC-Nov & 93.1M & 1.9B & 28-Oct-12 & 6\\\hline
\end{tabular}
\caption{Main characteristics of LCC snapshots}
\label{tab:lcc-dataset}
\vspace{-0.2cm}
\end{table}
\begin{table}
\small
\centering
\begin{tabular}{|c||c|c|c|c|c|}\hline
Name & \#nodes &  \#edges & Start Date & Duration (days)\\ \hline\hline
Rand-Apr & 2.2M & 145M & 8-Apr-12 & 23 \\ \hline
Rand-Oct & 5.7M & 263M & 15-Oct-12 & 10\\ \hline
Rand-Nov & 3.5M & 157M & 28-Oct-12 & 13\\ \hline
\end{tabular}
\caption{Main characteristics of Random datsets}
\label{tab:random-datasets}
\vspace{-0.4cm}
\end{table}

\noindent \textbf{Sampling Random Users:}
Our goal is to collect random samples of G+ users for our analysis.  To our knowledge, none
of the prior studies on G+ achieved this goal. The sparse utilization of the extremely large ID space makes it infeasible to identify random users by generating random IDs. To cope with this challenging
problem, we leverage the search function of the G+ API to efficiently identify a large number of 
seemingly random users. The function provides a 
list of up to 1000 users whose name or surname matches a given input keyword.
Careful inspection of search results for a few surnames revealed that G+
appears to order the reported users based on their level of connectivity and activity, 
\ie~users with a higher connectivity or activity (that are likely to be more interesting)
are listed at the top of the result. Since searching for popular surnames most likely
results in more than 1000 users,  the reported users are biased samples.
To avoid  this bias, we selected a collection of 1.5K random 
American surnames from the US\footnote{US is the most represented country in G+ \cite{g+_imc_cha, schioberg2012tracing}. Furthermore, the high immigration level of US allows to find surnames from different geographical regions.} 2000 census \cite{us_census} with low to moderate popularity
and used the search function of the API to obtain matched G+ users. We consider the list of 
reported users only if it contains less than 1000 users. These users are assumed
to be random samples because G+ must report all matched users, and
there is no correlation between surname popularity and the connectivity (or activity)
of the corresponding users. Table \ref{tab:random-datasets} summarizes the main characteristics of our
random datasets. Note that the timing of each one of the random datasets is aligned
with a LCC dataset.

\begin{figure*}[t]
\begin{minipage}{1.20\columnwidth}
	\subfigure[Num Followers]{\includegraphics[width=0.5\textwidth]
	{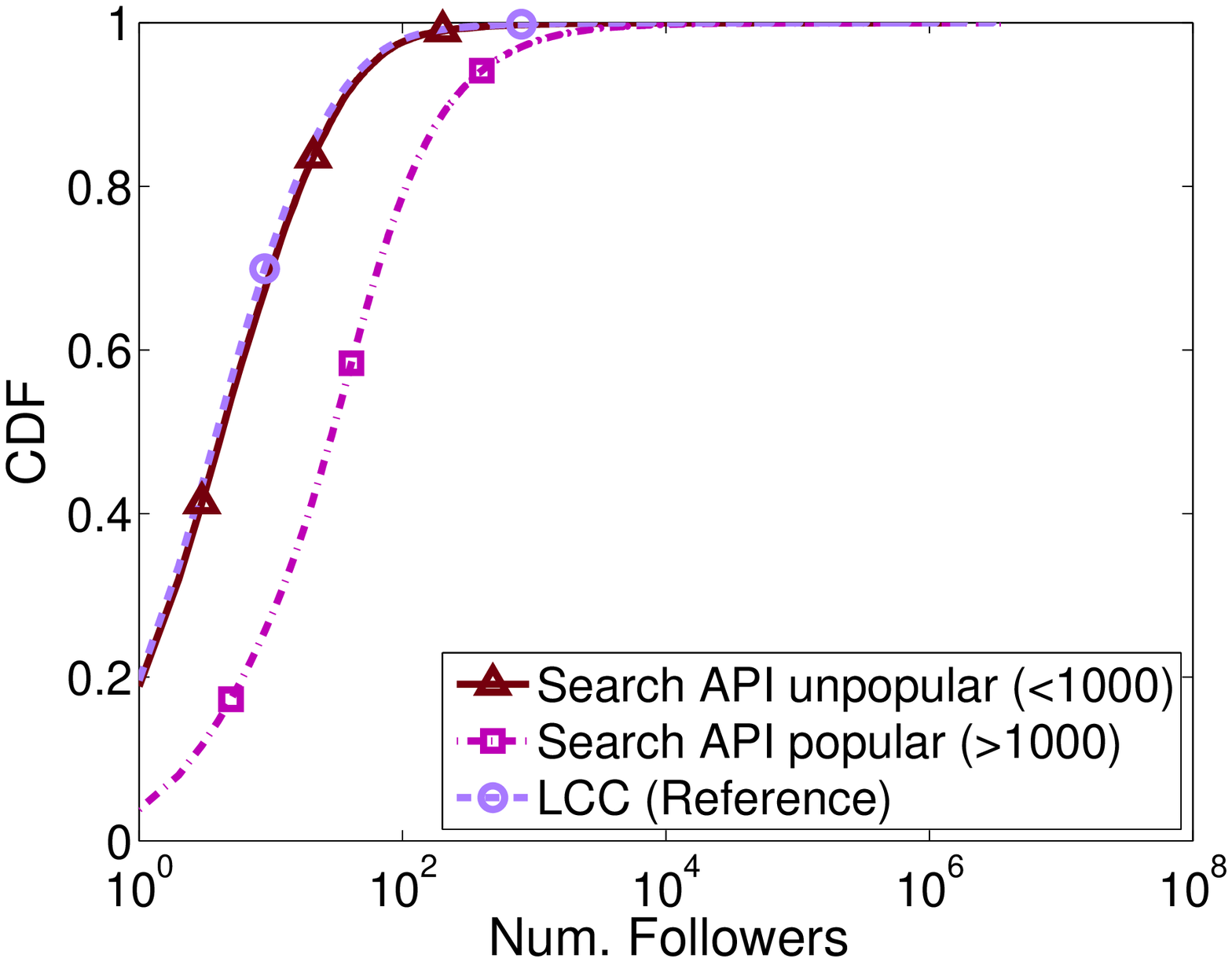}\label{fig:validation_foll}}\hfill
	\subfigure[Num Friends]{\includegraphics[width=0.5\textwidth]
	{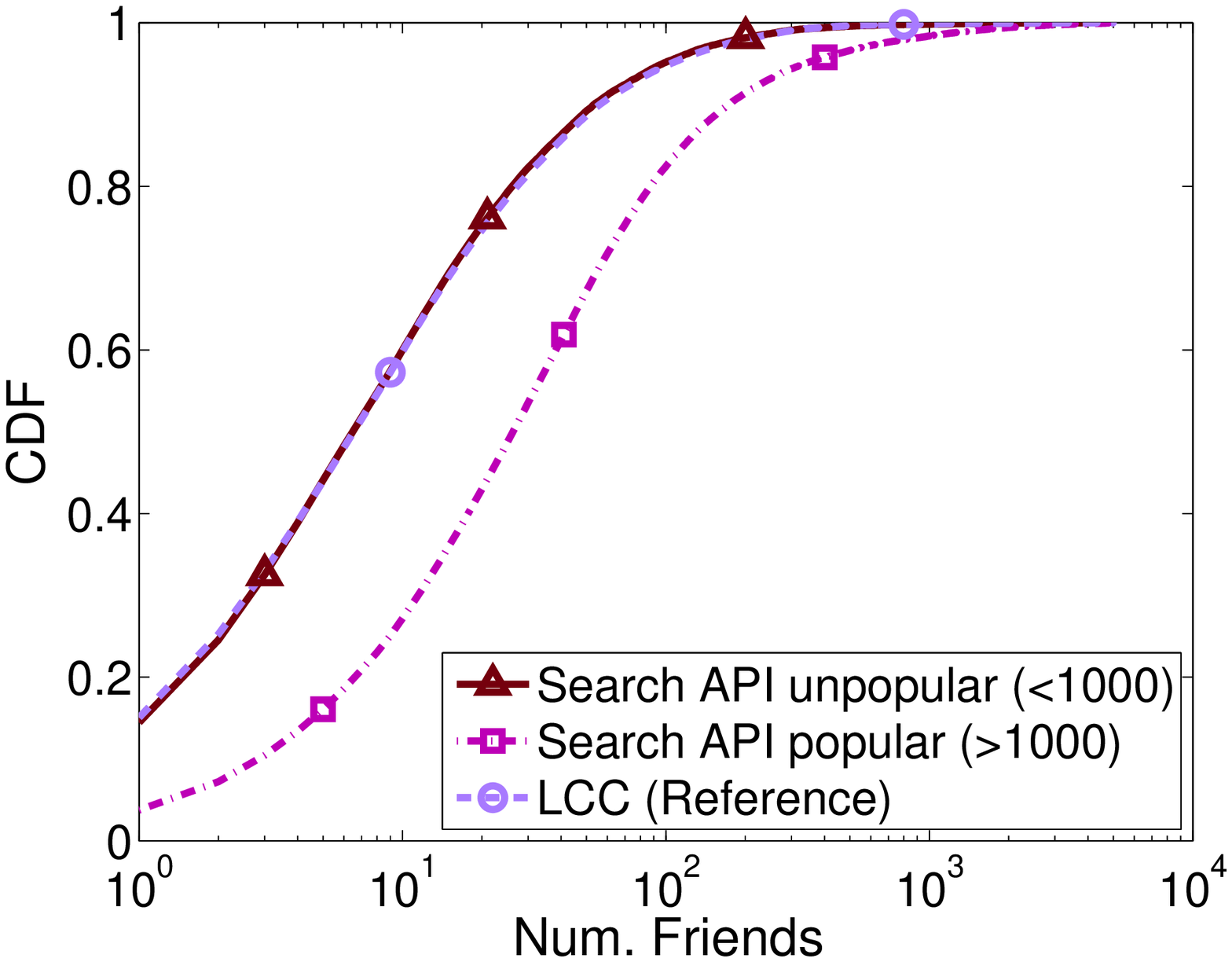}\label{fig:validation_friends}}
	\caption{Distribution of \#followers (a) and \#friends (b) for users collected from the search  function of G+ API with popular surnames ($>$1000 users), users collected with unpopular surnames ($<$ 1000 users), and all LCC users (Reference)}
	\label{fig:random_group}
\end{minipage}
\hfil
\begin{minipage}{0.80\columnwidth}
	\includegraphics[width=\textwidth,height=0.7\textwidth]	
	{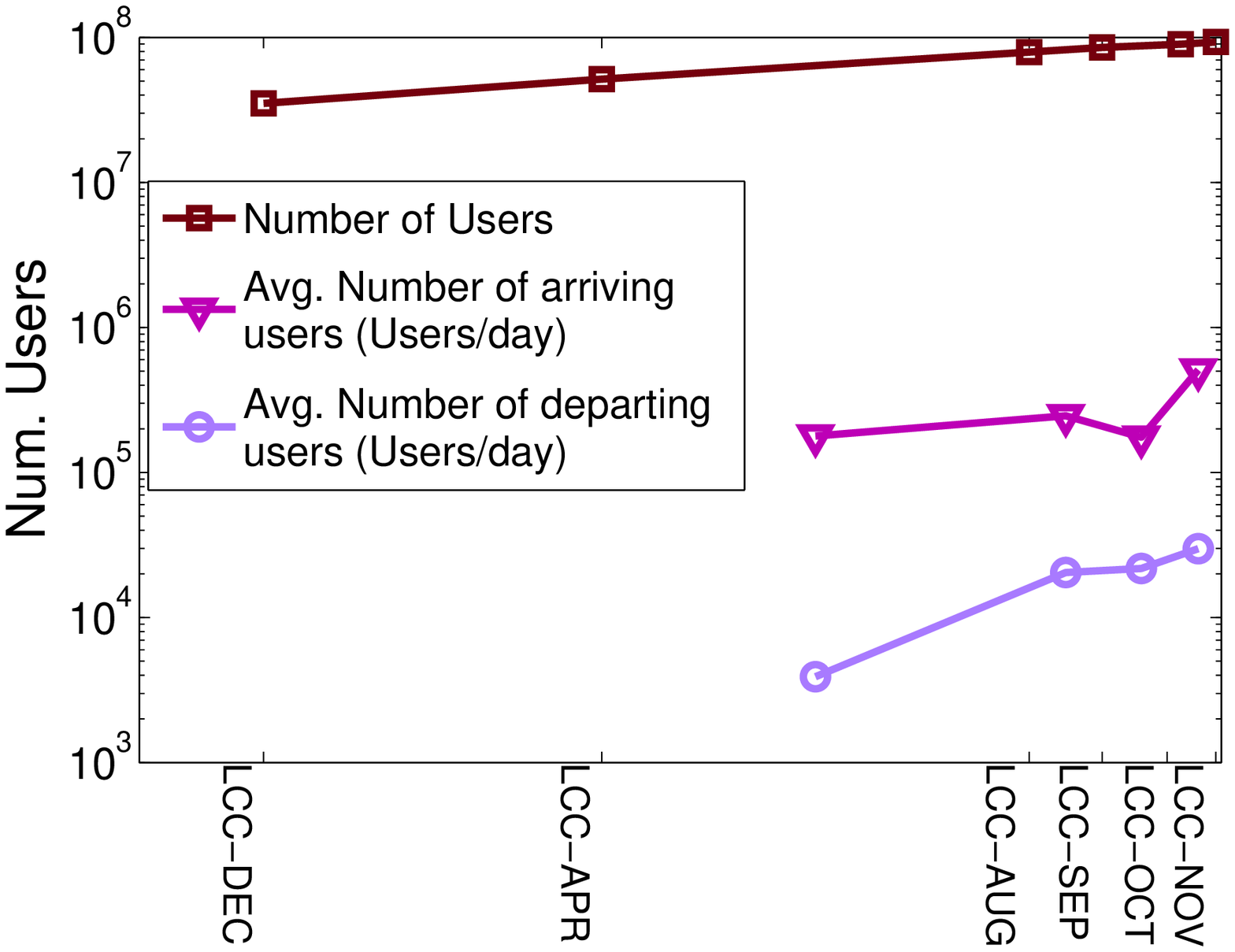}
	\caption{Evolution of total size and \#arriving and \#departing LCC users over time}
	\label{fig:growth}
\end{minipage}
\vspace{-0.3cm}
\end{figure*}

\begin{table}
\scriptsize
\centering
\hspace{-0.2cm}
\begin{tabular}{|c||c|c|c|c|c|}\hline
Users & Posts & Attachments & Plusones & Comments & Reshares\\\hline\hline 
13.6M & 218M & 299M & 352M & 202M & 64M\\ \hline 
\end{tabular}
\caption{Main characteristics of Activities among active users in the LCC (collected in Sep 2012)}
\label{tab:activities-datasets}
\vspace{-0.4cm}
\end{table}
\begin{table}
\scriptsize
\centering
\hspace{-0.2cm}
\begin{tabular}{|c||c|c|c|c|c|}\hline
Label & OSN & Date & Info\\\hline\hline 
TW-Pro & Twitter & Jul 2011 & Profile\\
& & & (80K rand. users) \\\hline
TW-Con \cite{cha2010measuring} & Twitter & Aug 2009 & Connectivity \\
& & & (55M users)\\\hline  
TW-Act \cite{Reza_IEEENetwork} & Twitter & Jun 2010 & Activity\\
& & & (895K rand. users)\\\hline  
FB-Pro & Facebook & Jun 2012 & Profile \\
& & & (480K rand. users) \\\hline
FB-Con & Facebook & Jun 2012 & Connectivity\\
& & &  (75K rand. users)\\\hline  
FB-Act & Facebook & Sep 2012 & Activity\\
& & & (16k rand. users)\\\hline  
\end{tabular}
\caption{Features of other  datasets in our analysis}
\label{tab:other_datasets}
\vspace{-0.6cm}
\end{table}
To validate the above strategy, we collect two groups of more than 140K  samples from the search API, users whose name match popular and unpopular ($<$ 1000 users) surnames, in Sep 2012. We focus on samples from each group that are located in the LCC since we have a complete snapshot of the LCC that can be used as ground truth. In particular, we compare the connectivity of samples from each group that are 
located in the LCC with all users in the LCC-Sep snapshot. Figure \ref{fig:random_group} plots the distribution of the number of followers and friends for these two groups of samples and all users in the LCC, respectively. These figures clearly demonstrate that only the collected LCC samples from unpopular surnames exhibit very similar distributions of followers and friends with the entire LCC. A Kolgomorov-Smirnov test confirms that they are indeed the same distribution. The collected samples from popular surnames have a stronger connectivity and thus are biased.

\noindent \textbf{Capturing User Activity:}
We consider user activity as a collection of all posts by individual users and the reaction (\ie~Plusones, Comments and Reshares) from other users to these posts.  User activity is an important indicator of user interest and thus the aggregate activity (and reactions) across users is a good measure of an OSN popularity. Despite its importance, we are not aware of any prior study that examined this issue among G+ users. Toward this end, we focus on user activity in the most important element of the network (\ie~the LCC). We leverage the G+ API to collect all the public posts and their associated reactions for  all LCC-Sep users between G+ release date (Jun 28$^{th}$ 2011) and the date our measurement campaign started (Sep 7$^{th}$ 2012), \ie~437 days. Given the cumulative nature of recorded activity for each user, a single snapshot of activity contains all the activities until our data collection time. Furthermore, since each post has a timestamp, we are able to determine the temporal pattern of all posts from all users. Note, that the G+ API limits the number of daily queries to 10K per registered application. Then, we use  303 accounts to collect the referred data in 68 days.   Table \ref{tab:activities-datasets} summarizes the main features of the activity dataset. In particular, note that only 13.6M (out of 85M) LCC-Sep users made at least one public post in the analyzed period.

\noindent \textbf{Other datasets:}
There are a few other datasets for Twitter and Facebook that we have either collected or obtained from other researchers. Table \ref{tab:other_datasets} summarizes the main features of these datasets. In the absence of any public dataset for Facebook, we developed our own crawler and collected the profile (FB-Pro) connectivity (FB-Con) and activity (FB-Act) for random Facebook users. We also collect the profile (TW-Pro) for random Twitter users.

%% file: elements-rr2.tex
\begin{table*}
\scriptsize
\centering
\begin{tabular}{|c||c|c|c||c|c|c||c|c|c|}\hline
	Element  &  \multicolumn{3}{c||}{\% users}&\multicolumn{3}{c||}{\% users } &\multicolumn{3}{c|}{\% users}   \\ 
	  &  \multicolumn{3}{c||}{}&\multicolumn{3}{c||}{public posts} &\multicolumn{3}{c|}{public attr.}   \\\hline\hline
	& Apr & Oct & Nov & Apr &  Oct & Nov & Apr &   Oct & Nov\\\hline
	LCC        & 43.5  & 32.3 & 32.2  & 8.9  & 7.0  & 6.9  & 27.4 &  17.9 & 17.6   \\ 
	Partitions & 1.4   & 1.7  & 1.5   & 0.1  & 0.2  & 0.2  & 0.5   & 0.6  & 0.5  \\
	Singletons & 55.1  & 66.0 & 66.3  & 1.4  & 1.6  & 1.6  & 1.8   & 5.7  & 6.2  \\\hline 
	All        & 100   & 100  & 100   & 10.4 & 8.8 &  8.7  & 29.7  & 24.2 & 24.3  \\\hline
	\end{tabular}
	\caption{Fraction of G+ users,  active users and  users with public attributes across G+ components along with the evolution of these characteristics from April to November of 2012 (based on the corresponding Random datasets)}
\label{tab:groups}
\vspace{-0.4cm}
\end{table*}

\section{Macro-Level Structure \& Its Evolution}
\label{sec:elements}
The macro-level connectivity structure among G+ users should intuitively consist of three components: $(i)$ The largest connected component (LCC), $(ii)$ A number of partitions  that are smaller than the LCC (with at least 2 users), and $(iii)$ Singletons or isolated users. We first examine the temporal evolution of the LCC size and then discuss the relative size of different components and their evolution over time.

\noindent
{\bf Evolution of the LCC Size:}
Having multiple snapshots of the LCC at different times enables us to examine the growth in the number of LCC users over time and determine the number of users who depart or arrive between two consecutive snapshots
as shown in Figure \ref{fig:growth} using log scale for the y axis. This figure illustrates that the overall size of the LCC has increased from 35M to 52M in four months between December 2011 and April 2012 at an average growth rate of 155K users per day. This average rate has even increased to 207K users per day between April and November 2012. 
The connectivity of these users to the LCC is a clear sign that they have intentionally joined  G+ by making the explicit effort to connect to other users (\ie,~these are interested users). While the daily rate of increase in the number of interested users (150K-200K) is impressive, it is an order of magnitude smaller than the .95-1.8M daily increase in the total population of G+ users that is officially reported by Google \cite{G+_official_stats}. The difference between the rate of growth for the overall system and LCC must be due to other components of the network (small partitions and singletons) as we explore  later in this section.

We observed some short term variations in the growth rate of LCC users (as shown in Figure \ref{fig:growth}) which is consistent with the reported results by another recent study on another large OSN \cite{gaito2012bursty}.
Figure \ref{fig:growth} also shows that LCC users have been departing the LCC at an average rate of  9.6K users per day. We carefully examined these departing users and discovered two points: {\em (i)} all of the departing users have removed their G+ accounts, and {\em (ii)} the distribution of \#followers, \#friends and public attributes  of departing users is very similar to all LCC users, however most of them are inactive. This seems to suggest that the departing users have lost their interest due to the lack of incentives to actively participate in the system.

\noindent
{\bf Evolution of the Main Components:}
To estimate the relative size of each component and its evolution over time, we determine the mapping of users in a random dataset to the three main components of the G+ structure.
LCC users can be easily detected using the corresponding LCC snapshot for each random dataset (\eg,~LCC-Oct for Rand-Oct). For all the users outside the LCC, we perform a BFS crawl from each user to verify whether a user is a singleton or part of a partition, and in the latter case determine the size of the partition.
The first part of Table \ref{tab:groups} presents the relative size of all three components using our random datasets in April, October and November of 2012\footnote{It is possible that our approach incorrectly categorizes user $u$ as a singleton if $u$ has a private list of friends and followers and, all of $u$'s friends and followers also have a private list of followers and friends. However, we believe this is rather unlikely. Indeed our BFS crawl on the LCC identified about 7.5\% users with private friend and follower lists who were detected through their neighbors.}. 
Table \ref{tab:groups} shows that the relative size of the LCC has dropped from 43\% (in Apr) to 32\% (in Oct and Nov) while the relative size of singletons has increased from 55\% to 66\% during the same period. Note that this drop in the relative size of the LCC occurs despite the dramatic increase in the absolute size of the LCC (as we reported earlier). This simply indicates an even more significant increase in the number of singletons. We believe that this huge increase in the number of singletons is a side effect of the integrated registration procedure that Google has implemented. In this procedure, a new G+ account is implicitly created for any user that creates a new Google account to utilize a specific Google service such as Gmail or YouTube\footnote{In fact, we examined and confirmed this hypothesis for new Gmail and YouTube accounts.}. The implicit addition of these new users to G+ suggests that they may not even be aware of (or do not have any interest in) their G+ accounts.
The relatively small and decreasing size of the LCC for the G+ network exhibits a completely different characteristic than the one reported for the LCC of other major OSNs. For instance, 99.91\% of the registered Facebook users were part of the LCC as of May 2011 \cite{Facebook1} and the LCC of Twitter included 94.8\% of the users with just 0.2\% singletons in August 2009 \cite{cha2010measuring}. Furthermore, Leskovec et al. \cite{leskovec2005graphs} showed that the relative size of the LCC of other social networks (\eg,~ the arXiv citation graph or an affiliation network) has typically increased with time until it included more than 90\% of their users.\\
Partitions make up only a small and rather stable fraction (1.5\%) of all G+ users.
We identified tens of thousands of such partitions and discovered that 99\% of these partitions have less than 4 users in all snapshots. The largest partition was detected in Rand-Apr snapshot with 52 users.

The last two parts of Table \ref{tab:groups} present the fraction of all G+ users that have at least one public post or provide public attributes in their profiles (on the last row) and the breakdown of these two groups across different components of the G+ network. We observe that the fraction of users that generate at least one public post has dropped from 10\% to 8.7\%, and the majority of them are part of the LCC. Similarly, the fraction of users with at least one public attribute have dropped from roughly 30\% to 24\% over the same period. A large but decreasing fraction of these users are part of the LCC and a smaller but growing fraction of them are singletons.
Since the LCC is the well connected component that contains the majority of active users, we focus our remaining analysis only on the LCC.

{\em In summary, the absolute size of the LCC in the G+ network has been growing by 150-200K users/day while its relative size has been decreasing. This is primarily due to the huge increase in the number of singletons that is caused by the implicit addition of new Google account holders to G+. In November of 2012, the LCC made up 1/3rd and the rest of the network mostly consists of singletons. Less than 9\% of G+ users generate any post, and less than 25\% provide any public attribute, and a majority of both groups are LCC users}.

%% file: activity-rr2.tex
\begin{figure*}[t]
\centering
	\subfigure[The num. of daily posts, and num. of daily posts with different types of reactions]{\includegraphics[width=0.32\textwidth,height=0.20\textwidth]{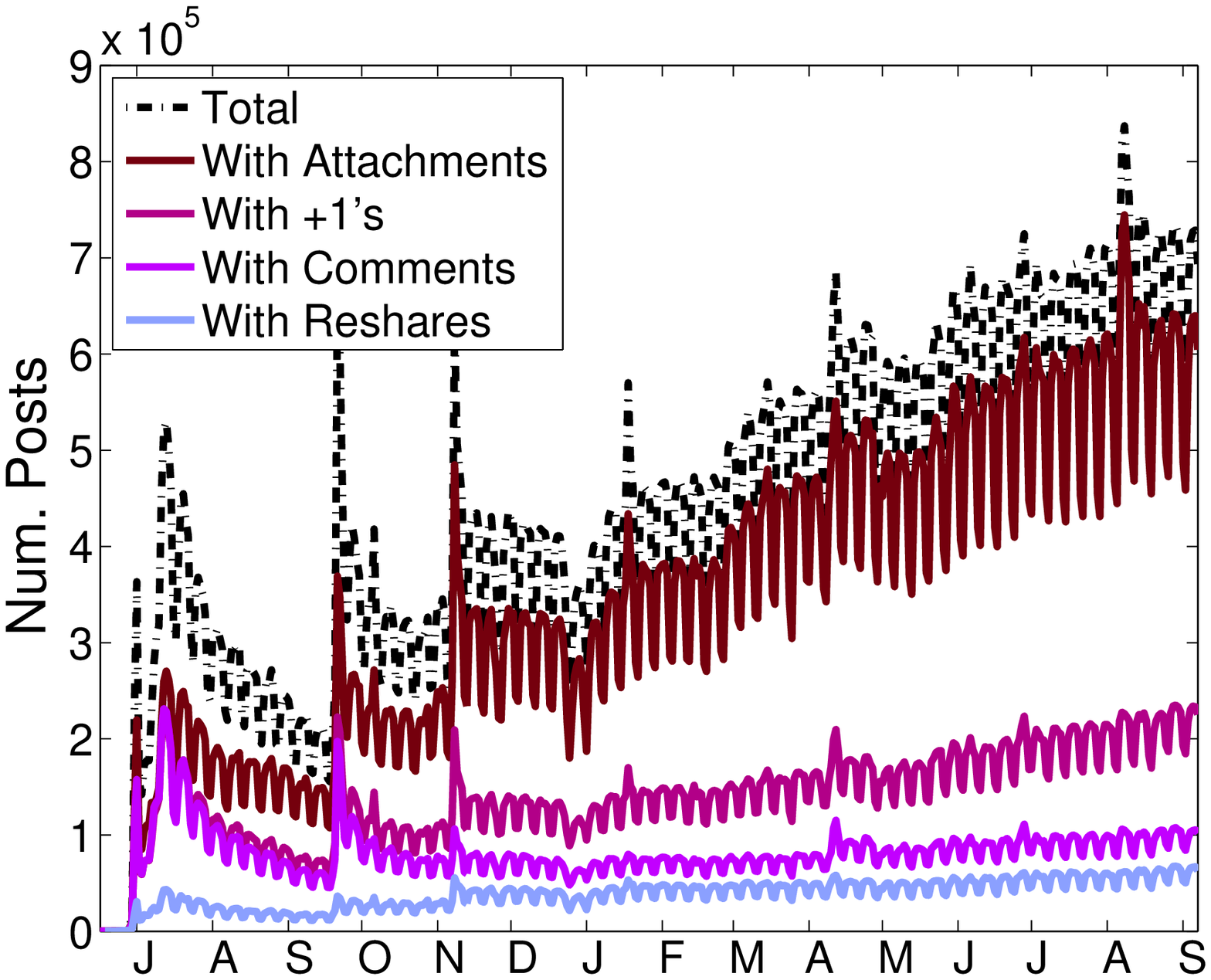}\label{fig:posts_and_reactions}}\hfill
	\subfigure[The num. of daily attachments, plusones, reshares and comments.]{\includegraphics[width=0.32\textwidth,height=0.20\textwidth]{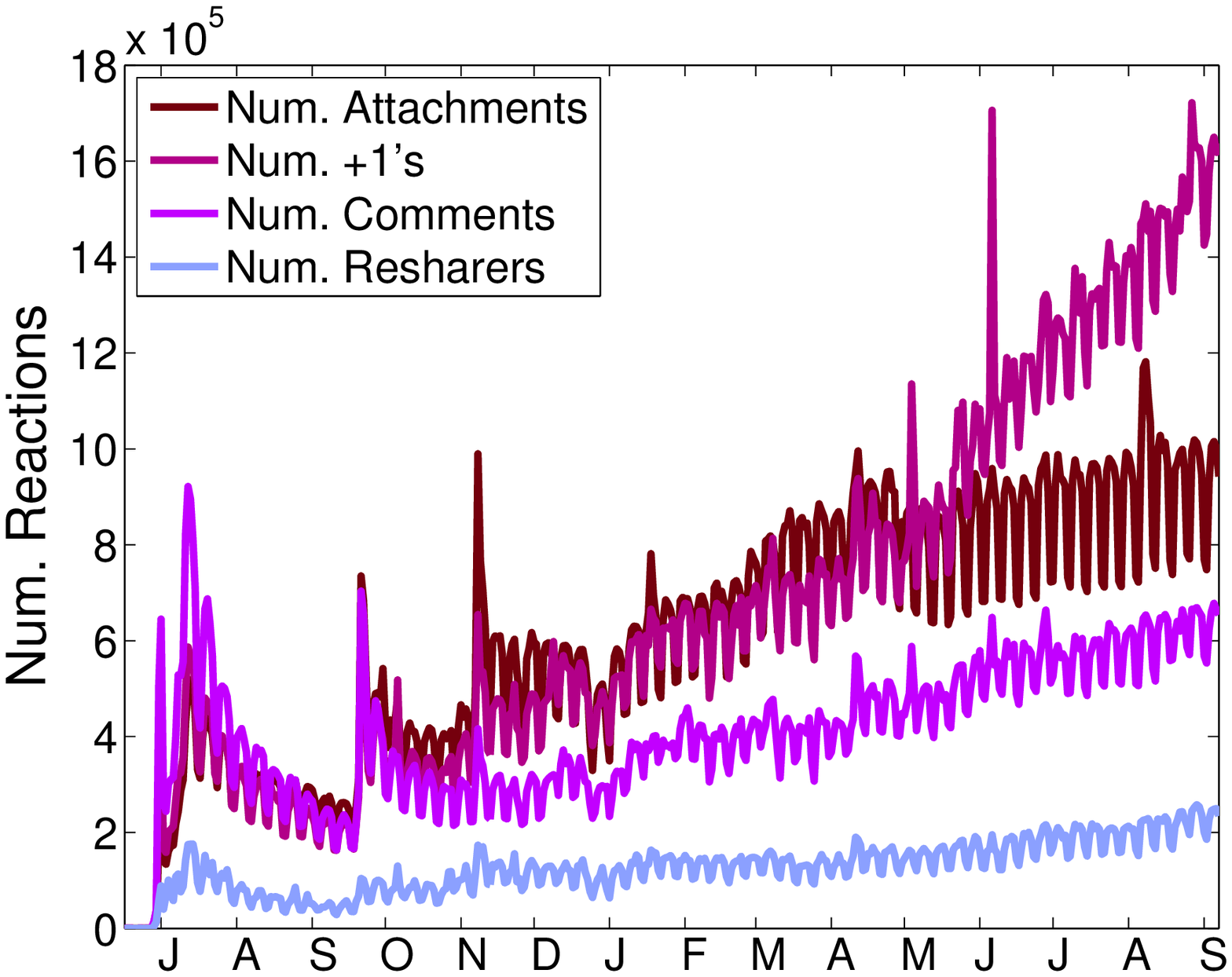}\label{fig:reactions}}\hfill
	\subfigure[The num. of daily active users and num. of daily active users whose posts received each type of reaction]{\includegraphics[width=0.32\textwidth,height=0.20\textwidth]{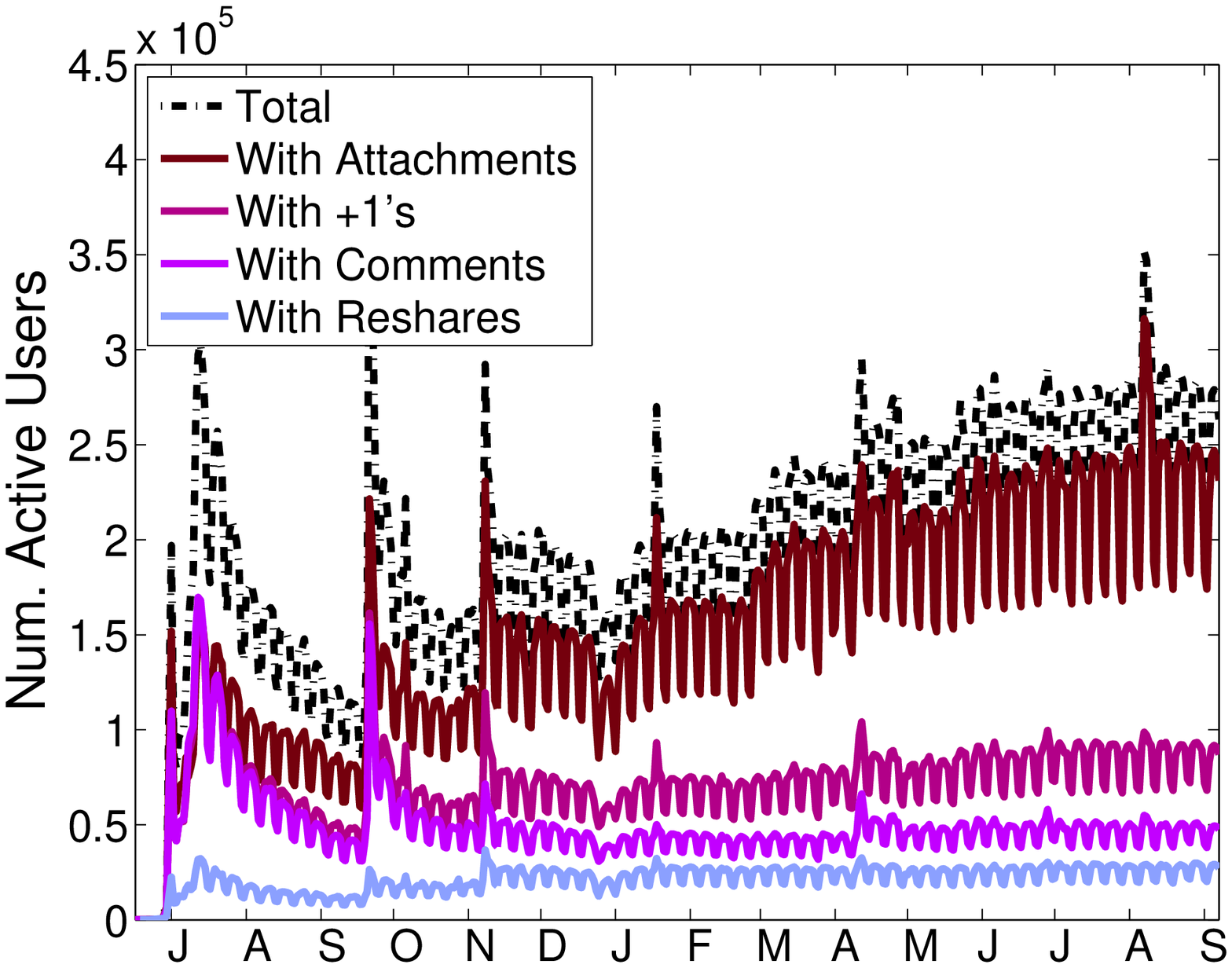}\label{fig:users_and_reactions}}
	\vspace{-0.2cm}
	\caption{Evolution of different aspects of public user activity during the 15 months operation of G+ (July 2011 to September 2012)}
	\label{fig:activity}
	\vspace{-0.5cm}
\end{figure*}

\section{Public Activity \& Its Evolution}
\label{sec:act}
To investigate users activity, we characterize publicly visible (or in short "public") posts by LCC users as well as other users' reactions (including users outside the LCC) to these public posts\footnote{We are not aware of any technique to capture private posts in G+ for obvious reasons. It might be feasible to create a G+ account and connect to a (potentially) large number of users in order to collect their private posts. However, such a technique is neither representative nor ethical.}. 

An earlier study used ground-truth data to show that more than 30\% of posts in G+ were public during the initial phase of the system \cite{kairam2012talking}.
However, the proposed setting by Google encourages users to generate public posts and reactions since only these public activities are indexable by search engines (including Google), and thus visible to others (apart from Google) for various marketing and mining purposes \cite{google_privacy}. Therefore, characterizing public posts and their reactions provides an important insight about the publicly visible part of G+.

We recall that the main {\em action} by individual users is to generate a ``post'' that may have one or more ``attachments''. Each post by a user may trigger other users to react by making a ``comment'', indicate their interest by a ``plusone'' (+1) or ``reshare'' the post with their own followers. To maintain the desired crawling speed for collecting activity information, we decided to only collect the timestamp for individual posts (but not for reactions to each post). Therefore, we use the timestamp of each post as a good estimate for all of its reactions because most reactions often occur within a short time after the initial post. To validate this assumption, we have examined the timestamp of 4M comments associated to 700K posts and observed that more than 80\% of the comments occurred within the 24 hours after their corresponding post. 

\noindent \textbf{Temporal Characteristics of Public Activity:}
Having the timestamp for all the posts and their associated reactions enables us to examine the temporal characteristics of all public activity among LCC users during the entire 15 months of G+ operation until our measurement campaign started.

Figure \ref{fig:posts_and_reactions} depicts the total number of daily posts by LCC users along with the number of daily posts that have attachments, have at least one plusone, have been reshared or have received comments.
Note that a post may have any combination of attachments, plusones, reshares and comments (\ie,~these events are not mutually exclusive). The pronounced repeating pattern in this figure (and other similar results) is due to the weekly change in the level of activity among G+ users that is significantly lower  during the weekend and much higher  during weekdays.
The timing of most of the observed peaks in this (and other related) figure(s) appears to be perfectly aligned with specific events as follows\footnote{We could not identify any significant event at the time of the peaks on May 3rd, Jun 4th and Aug 7th.}: 
{\em (i)} the peak on Jun 30th caused by the initial release of the system (by invitation) \cite{G+_wiki}; 
{\em (ii)} the peak on Jul 11th is due to users reaction to a major failure on Jul 9th when the system ran out of disk \cite{apology_g+};
{\em (iii)} the peak on Sep 20th was caused by the public release of the system \cite{G+_wiki};
{\em (iv)} the peak on Nov 7th is due to the release of G+ Pages service  \cite{gpages_announcement};
{\em (v)} the peak on Jan 17th is caused by the introduction of new functionalities for auto-complete and adding text in photos \cite{G+_new_features_2,G+_new_features_1};
{\em (vi)} on Apr 12th, caused by a major redesign of G+\cite{g+_redesign}.
Figure \ref{fig:posts_and_reactions} also demonstrates that the aggregate number of daily posts has steadily increased after the first five months (\ie,~the initial phase of operation). We can observe that a significant majority of the posts have attachments but the fraction of posts that trigger any reaction by other users is much smaller, in addition plusones is the most common type of reaction. Note that Figure \ref{fig:posts_and_reactions} presents the number of daily posts with attachments or reactions but does not reveal the total daily number of attachments or reactions. To this end, Figure \ref{fig:reactions} depicts the temporal pattern of the aggregate daily rate of attachments, plusones, comments and reshares for all the daily posts by LCC users, \ie,~multiple attachments or reactions to the same post are counted separately. This figure paints a rather different picture. More specifically, the total number of comments and in particular plusone reactions have been rapidly growing after the initial phase.
Figure \ref{fig:reactions} illustrates that individual posts are more likely to receive multiple plusones than any other type of reaction, and its comparison with Figure \ref{fig:posts_and_reactions} shows that most post have one or two attachments.
Figure \ref{fig:users_and_reactions} plots the temporal pattern of user-level activity by showing the daily number of active LCC users along with the number of users for whom their posts have attachments or triggered at least one type of reaction. This figure reveals that the total number of users with a public post has been steadily growing (after the initial phase) roughly at the rate of 3K users per day. However, this rate of growth in active users is significantly (roughly 60 times) lower than the rate of growth of LCC users which means only a very small fraction of new LCC users ($<$ 2\%) ever become active. 
While a large fraction of these users create posts with attachments, the number of daily users whose posts trigger at least one plusone, comment or reshare has consistently remained below 1M, 0.5M and 0.25M, respectively, despite the dramatic growth in the number of LCC users.
\begin{figure*}[t]
\begin{minipage}{1.33\columnwidth}
	\subfigure[\% of posts, attachments, plusones, reshares, comments associated to top x\% users]{\includegraphics[width=0.48\textwidth]
	{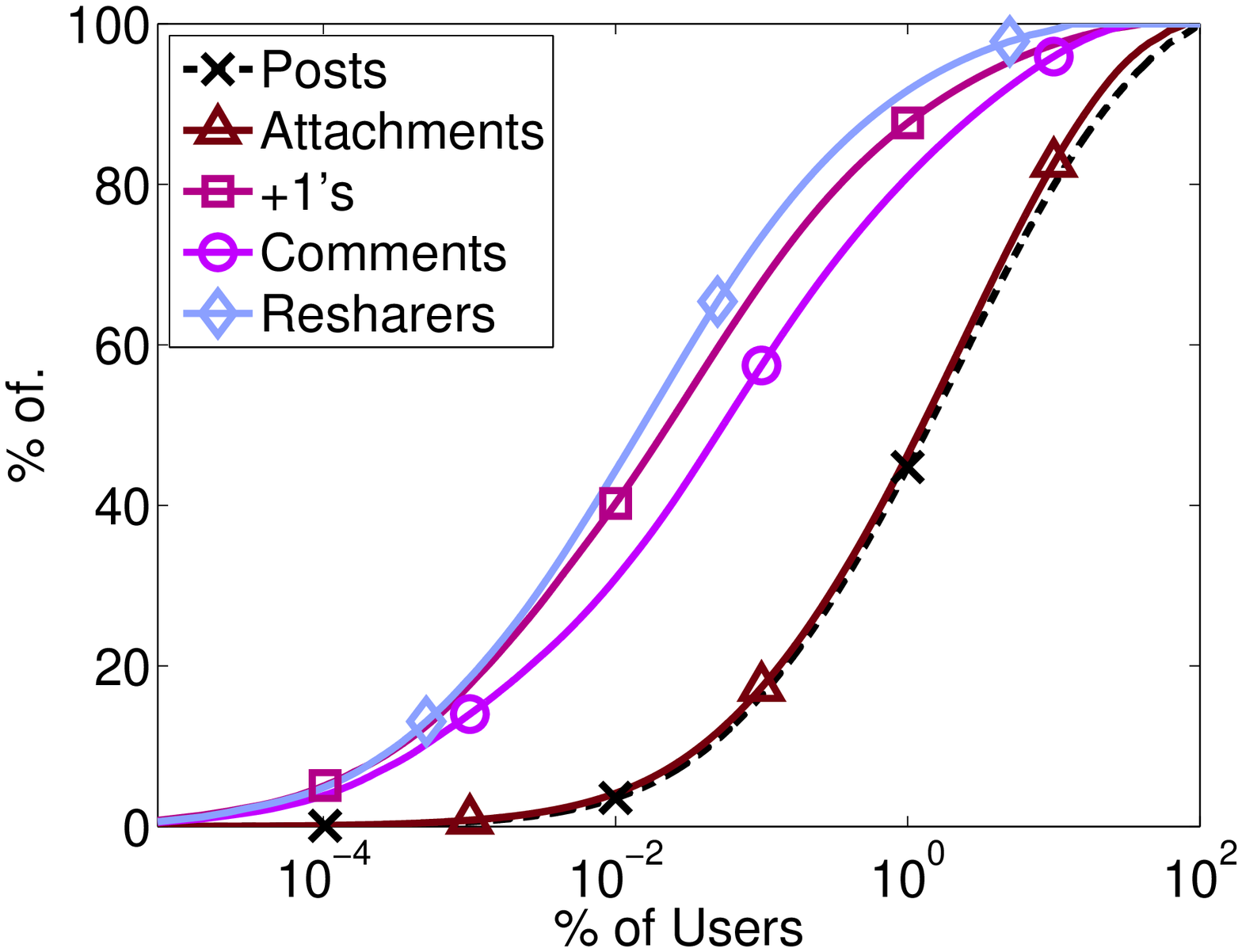}\label{fig:skewness_users}}
	\hfill
	\subfigure[\% of attachments, plusones, reshares, comments associated to top x\% posts]{\includegraphics[width=0.48\textwidth]
	{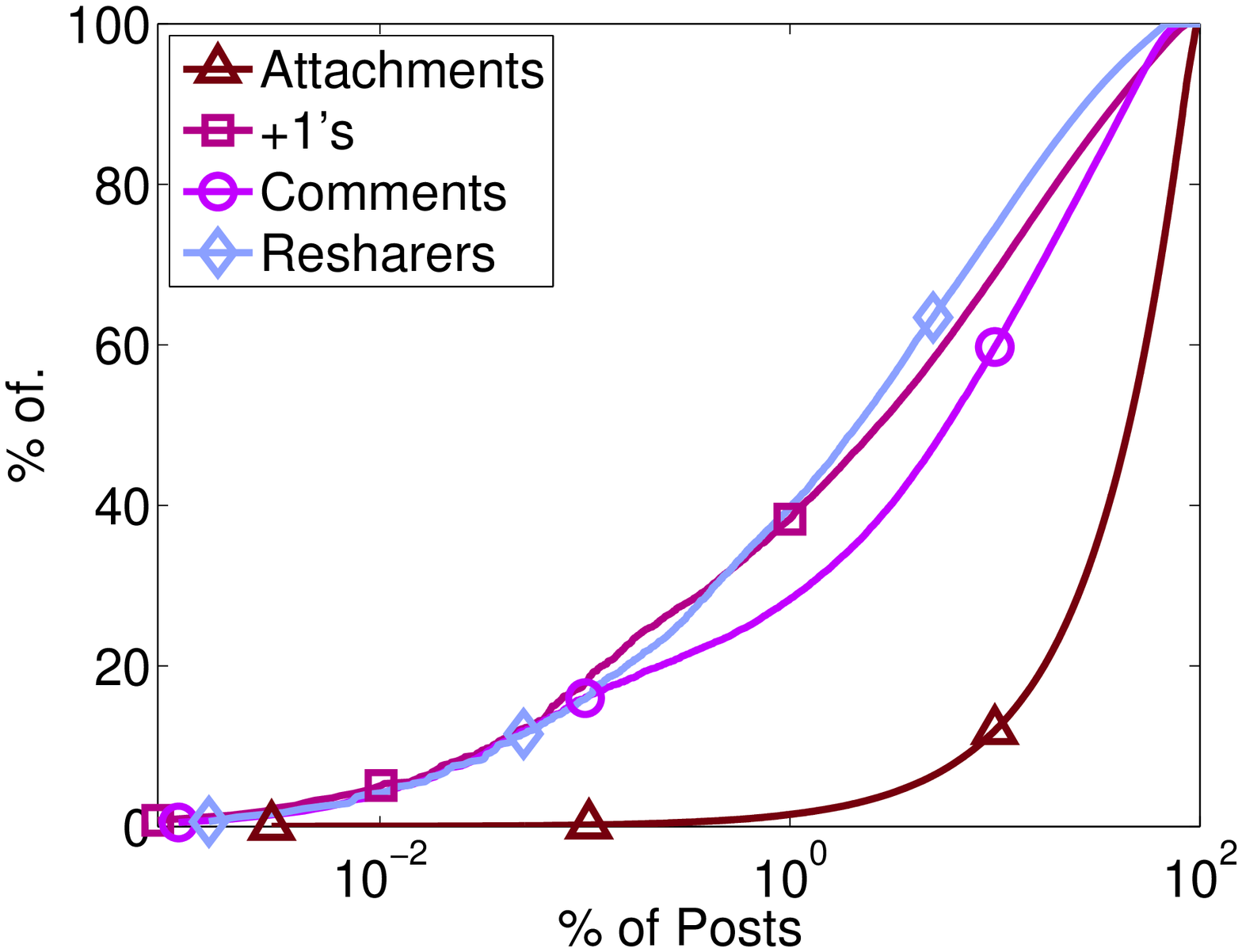}\label{fig:skewness_posts}}
	\vspace{-0.4cm}
	\caption{Skewness of actions and reactions contribution per user and post}
	\label{fig:skewness}
	
\end{minipage}
\hfill
\begin{minipage}{0.67\columnwidth}
\includegraphics[width=\textwidth]
{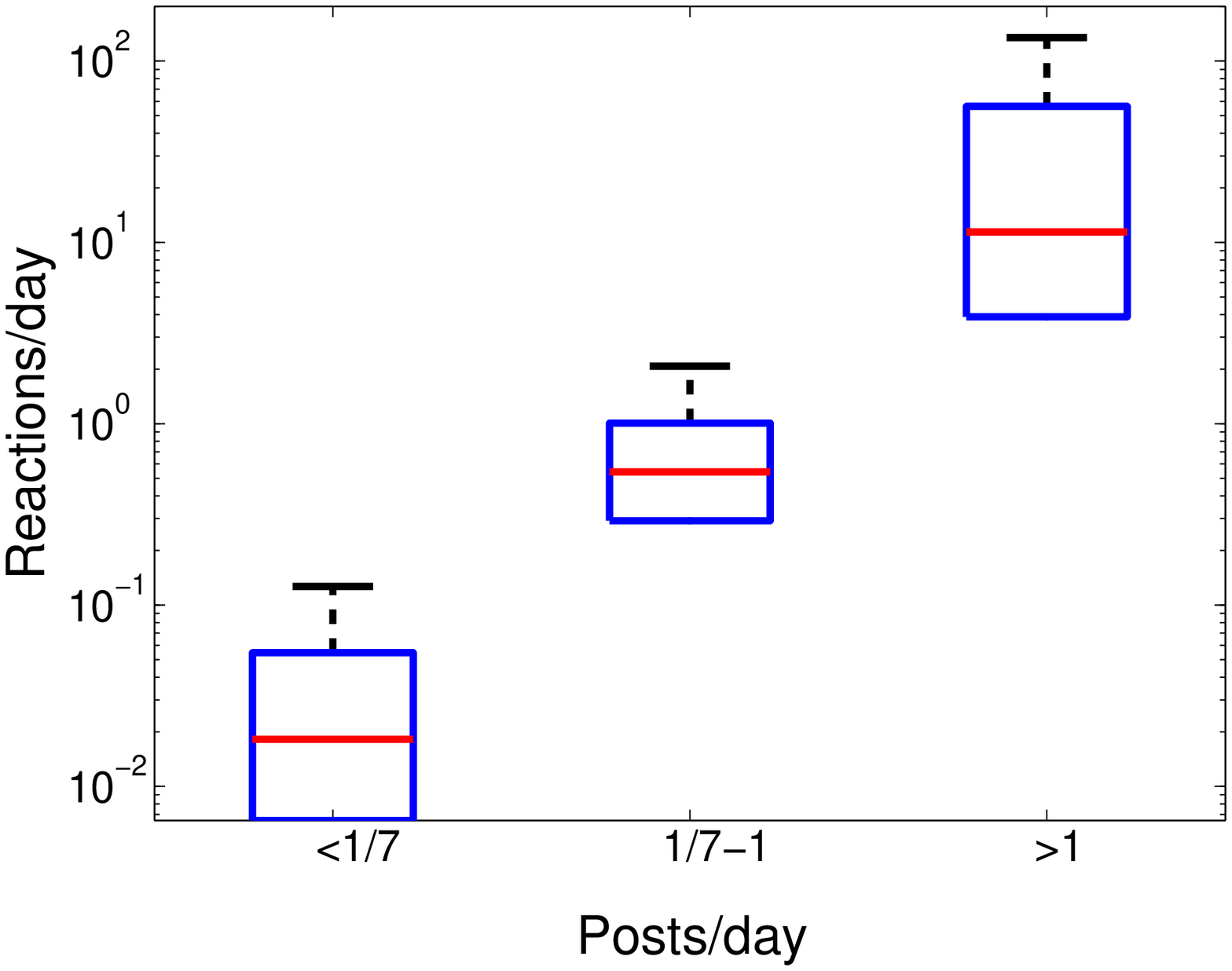}
\caption{Post-rate (x axis) vs aggregate reaction rate (y axis) correlation}
\label{fig:action_vs_reaction}
\end{minipage}
\vspace{-0.3cm}
\end{figure*}

\begin{figure*}[t]
\begin{minipage}{1.34\columnwidth}
	\centering
	\subfigure[Avg. Post/Tweet rate]{\includegraphics[width=0.5\textwidth]
	{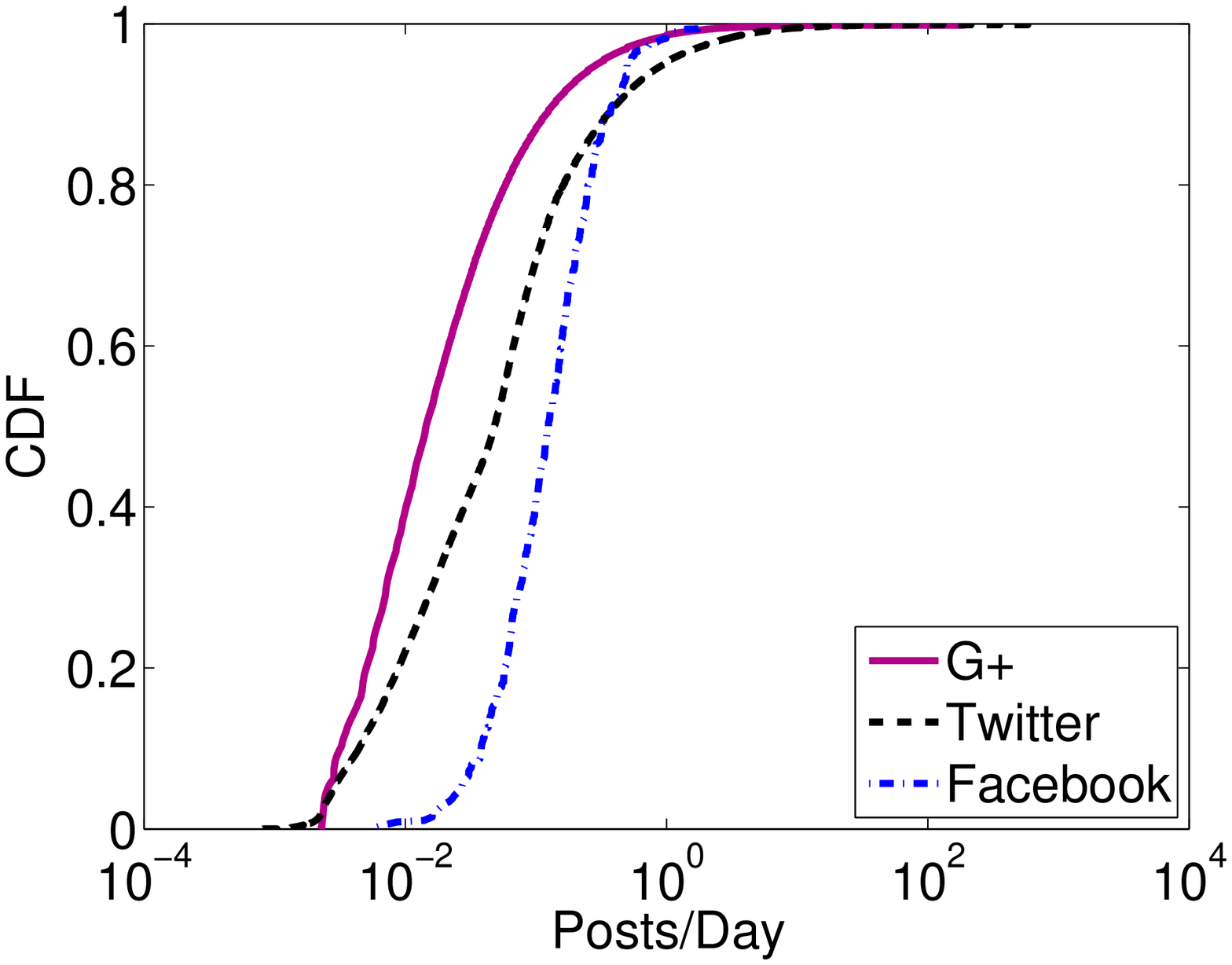}\label{fig:posting_rate}}\hfill
	\subfigure[Recency of Activity]{\includegraphics[width=0.5\textwidth]
	{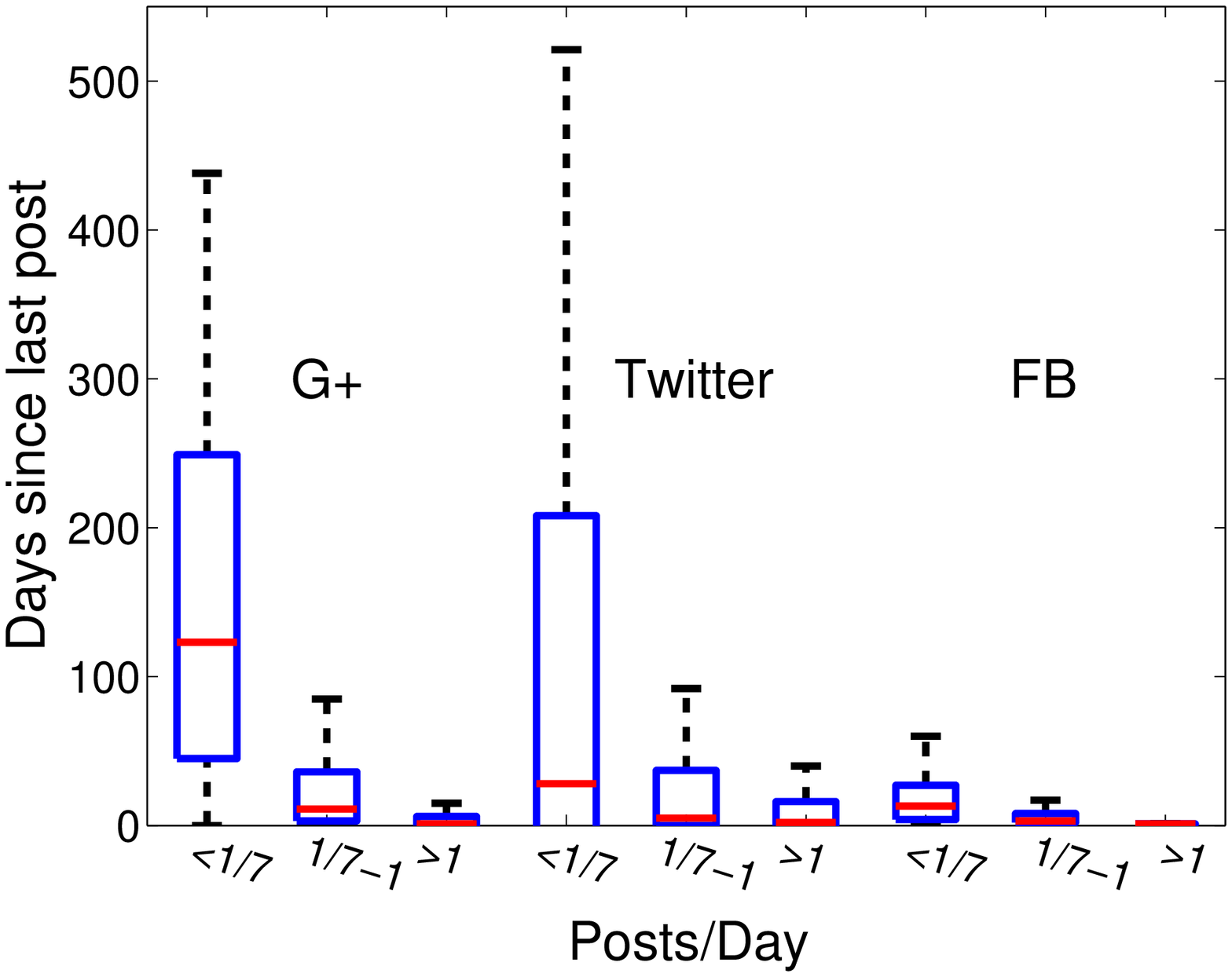}\label{fig:recency}}
	\caption{Comparison of activity metrics for G+, Twitter and Facebook}
	\label{fig:activity_comparison}
\end{minipage}
\hfill
	\begin{minipage}{0.67\columnwidth}
		\centering
		\includegraphics[width=\textwidth]
		{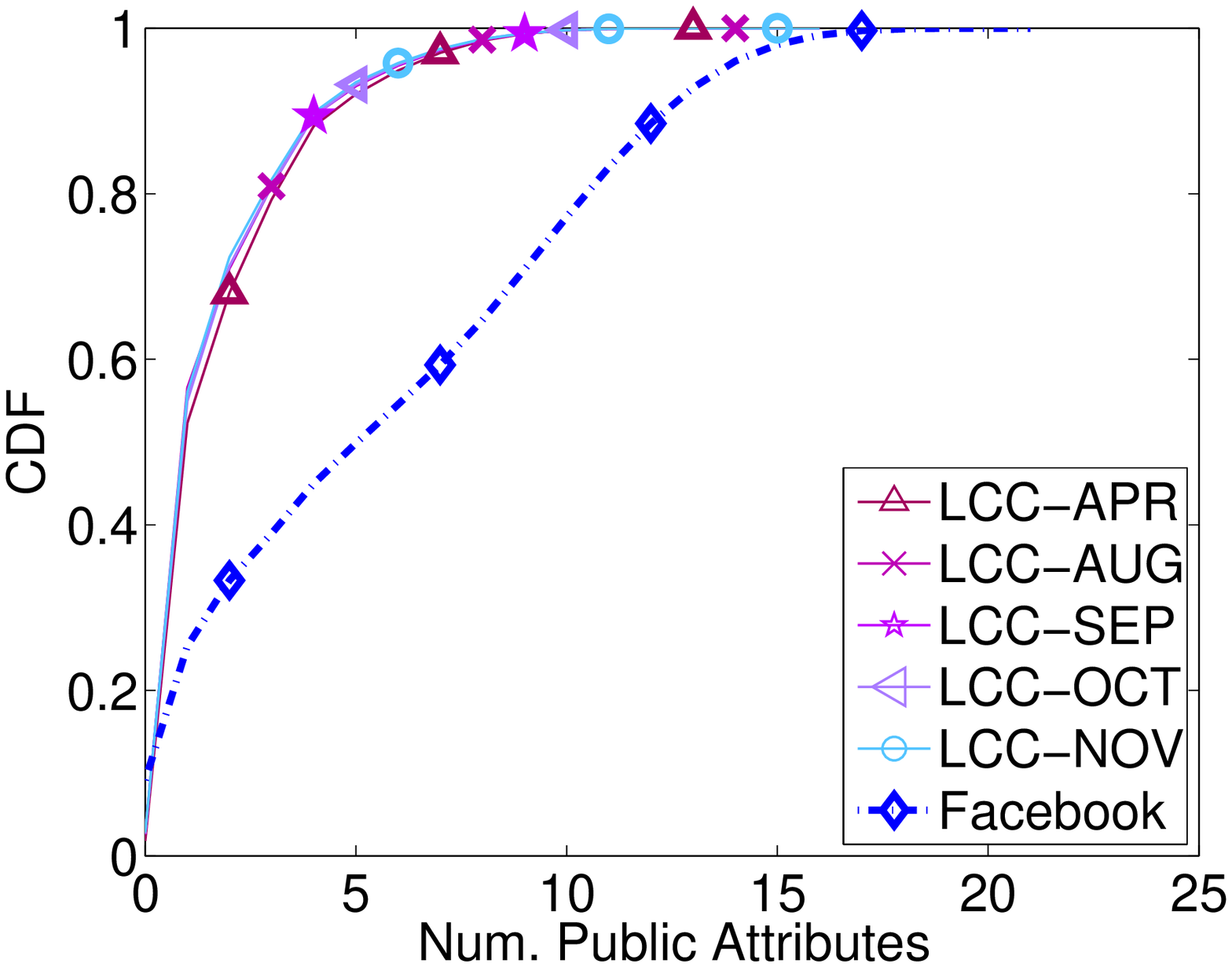}
		\caption{Distribution of number of public attributes for G+ and Facebook}
		\label{fig:cdf_public_attributes}
	\end{minipage}
\vspace{-0.5cm}	
\end{figure*}

\noindent \textbf{Skewness in Activity Contribution:}
We observed that a relatively small and stable number of users with interesting posts receive most reactions. This raises the question that ``how skewed are the distribution of generated posts and associated reactions among users in G+?''.  Figure \ref{fig:skewness_users} presents the fraction of all posts in our activity dataset that are generated by the top
$x$\% of LCC users during the life of G+ (the x axis has a log-scale). Other lines in this figure show the fraction of all attachments, plusones, comments and reshares that are associated with the top $x$\% users that receive most reactions of each type. This figure clearly demonstrates that the contribution of the number of posts and the total number of associated attachments across users is similarly very skewed. For example, the top 10\% of users contribute 80\% of posts. Furthermore, the distribution of contribution of received reactions to a user's posts is an order of magnitude more skewed than the contribution of total posts per user. In particular, 1\% of users receive roughly 80\% of comments and 90\% of plusones and reshares. These findings offer a strong evidence that {\em only a very small fraction of users (around 1M) create most posts and even a smaller fraction of these users receive most reactions from other users to their posts, \ie,~both user action and reaction are centered around a very small fraction of users.} We also repeated a similar analysis at the post level to assess how skewed are the number of reactions to individual posts. Figure \ref{fig:skewness_posts} shows the fraction of attachments, plusones, comments and reshares associated to the top $x$\% posts. The distribution for attachments is rather homogeneous which indicates that most posts have one or small number of attachments. For other types of reactions, the distribution is roughly an order of magnitude less skewed that the distribution of reaction across users (Figure \ref{fig:skewness_users}) .This is a rather expected result since reactions tend to spread across different posts by a user. 

\noindent \textbf{Correlation Between User Actions and Reactions:}
Our analysis so far has revealed that actions and reactions are concentrated on a small fraction of LCC users. However, it is not clear whether users who generate most of the posts are the same users who receive most of the reactions. For example, a celebrity may generate a post infrequently but receives lots of reaction to each post. To answer this question, first we examine the correlation between the number of posts and the aggregate reaction rate for different groups of users grouped based on their average level of activity as follows:\\
\noindent {\hspace{0.2cm}\em -Active} users who post at least once a day ($>$1),\\
\noindent {\hspace{0.2cm}\em -Regular} users who post less than once a day but more than once a week ($\frac{1}{7}$-1), and\\ 
\noindent {\hspace{0.2cm}\em -Casual} users who post less than once a week ($<\frac{1}{7}$).

Figure \ref{fig:action_vs_reaction} shows the summary distribution of daily reaction rate among users in each one of the described groups using boxplots. This figure reveals that the reaction rate grows exponentially with the user posting rate. 
Therefore, {\em the small group of users that contribute most posts is also receiving the major portion of all reactions. }

We have inspected the identity of the top 20 users with a largest number of public posts to learn more about them as well as those that receive a largest number of reactions. While the analysis of the first group does not reveal any interesting finding, we observe that 18 of the top 20 users attracting more reactions are related to music groups by young girls from Japan and Indonesia (\eg,~nmb48, ske48, akb48, hkt48 from Japan or jkt48 from Indonesia). All these groups are associated to the same Japanese record producer (Yasushi Akimoto) whose G+ account is also among the top 20.

\subsection{Comparison with Other OSNs}
We examine a few aspects of user activity (\ie,~generating posts or tweets) among G+, Twitter and Facebook users to compare the level of user engagement in these three OSNs. For this comparison, we leverage TW-Act and FB-Act datasets (described in Table \ref{tab:other_datasets}) that capture activity of random users in the corresponding OSNs. In our analysis, we only consider the active users in each OSN that make up 17\%, 35\%, and 73\% of all users in G+, Facebook and Twitter, respectively.

\noindent{\bf Activity Rate:} Figure \ref{fig:posting_rate} shows the distribution of the average activity rate per user across  all active users in each OSN. The activity rate is measured as the total number of posts or tweets divided by the time between the timestamp of a user's first collected action and our measurement time. This figure reveals the following two basic points in comparing these three OSNs: {\em (i)} the activity rate among Facebook and G+ users are more homogeneous than across Twitter users, {\em (ii)} Facebook users are the most active (with a typical rate of 0.19 posts/day) while G+ users exhibit the least activity rate (with a typical rate of 0.08 posts/day).



\noindent
{\bf Recency of Last Activity:} An important aspect of user engagement is how often individual users generate a post. We can compute the \emph{recency} of the last post by each active user as the time between the timestamp of the last post and our measurement time. The distribution of this metric across a large number of active users provides an insight on how often active users generate a post. Figure \ref{fig:recency} depicts the distribution of recency of the last post across G+, Twitter and Facebook users. We have divided the users from each OSN into three groups of casual, regular and active users based on their average activity rate ($<\frac{1}{7}$, $\frac{1}{7}$-1, $>$1 post/day) as we described earlier.
We observe that among casual users in all three OSNs, Facebook and Twitter users typically generate posts much more frequently (\ie,~have lower median recency) than casual G+ users. Regular users in different OSNs exhibit the same relative order in their typical recency of last post. Finally, for active users, it is not surprising to observe that all three OSNs show roughly the same level of recency.


\begin{figure*}[t]
\begin{minipage}{\columnwidth}
	\subfigure[Num Followers]{\includegraphics[width=0.5\textwidth]
	{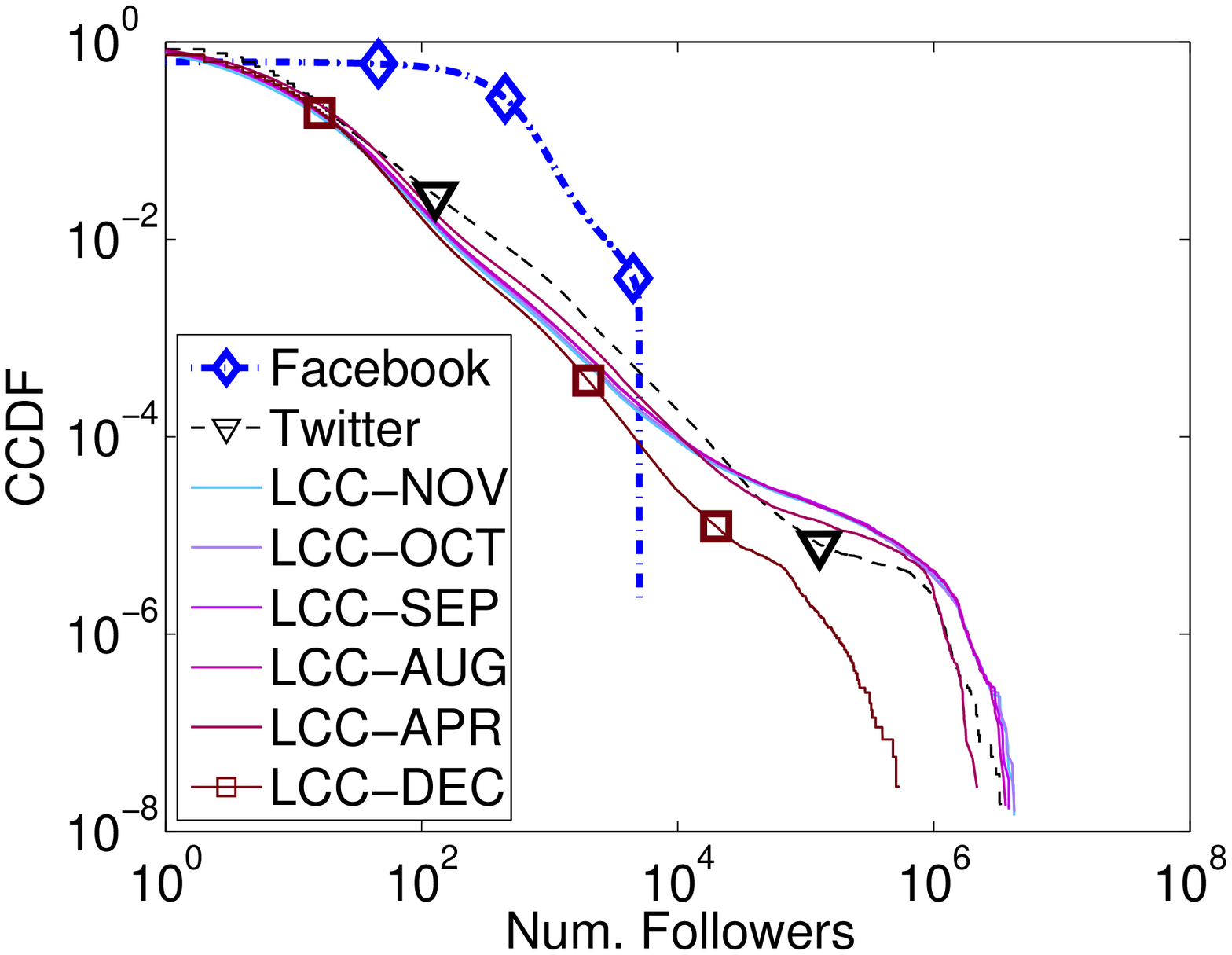}\label{fig:cdf_followers}}\hfill
	\subfigure[Num Friends]{\includegraphics[width=0.5\textwidth]
	{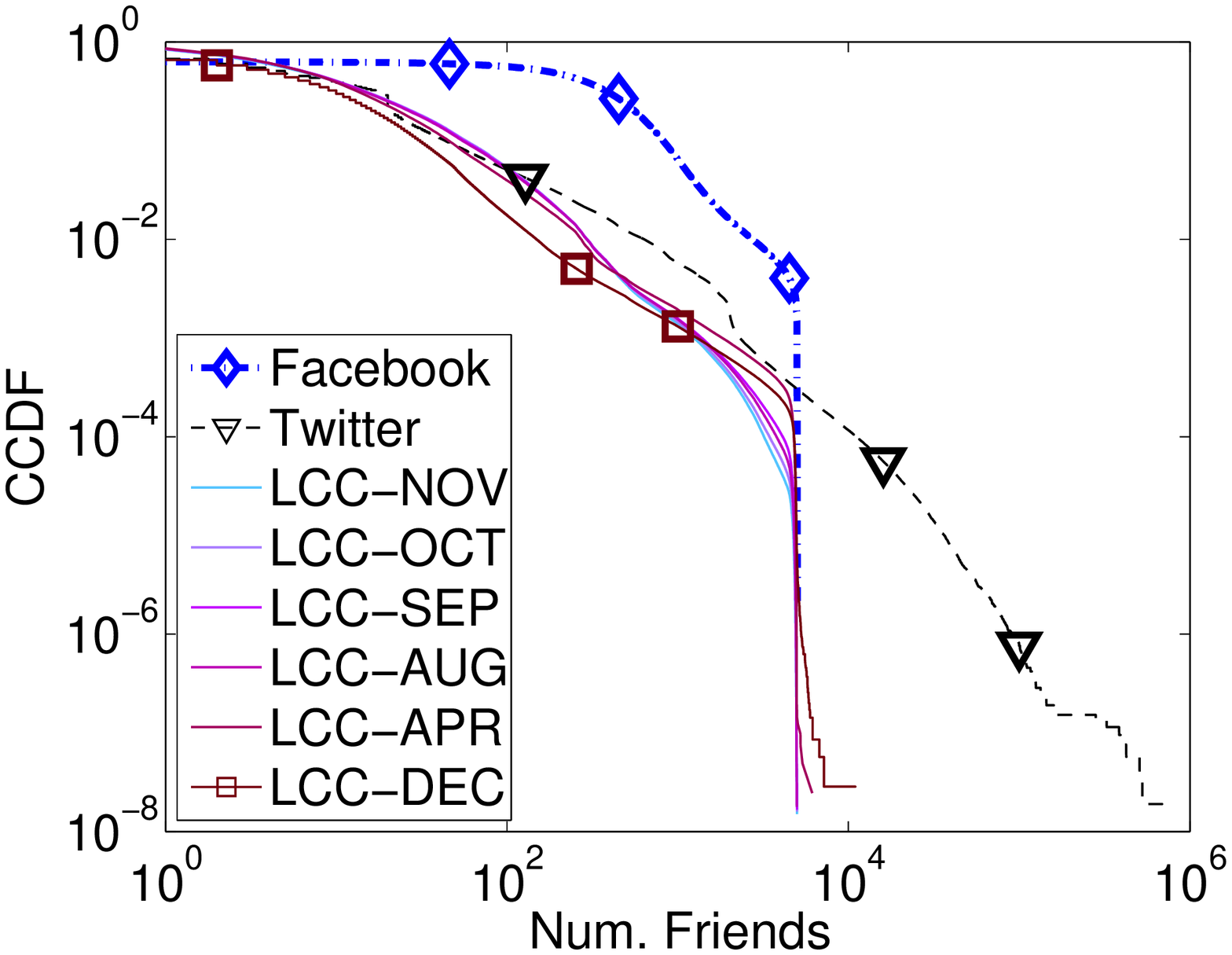}\label{fig:cdf_friends}}
	\caption{Degree Distribution for different snapshots of G+, Twitter and Facebook}
	\label{fig:degree_distribution}
\end{minipage}
\hfill
\begin{minipage}{\columnwidth}
	\centering
	\subfigure[\#followers/\#friends]{\includegraphics[width=0.5\textwidth]
	{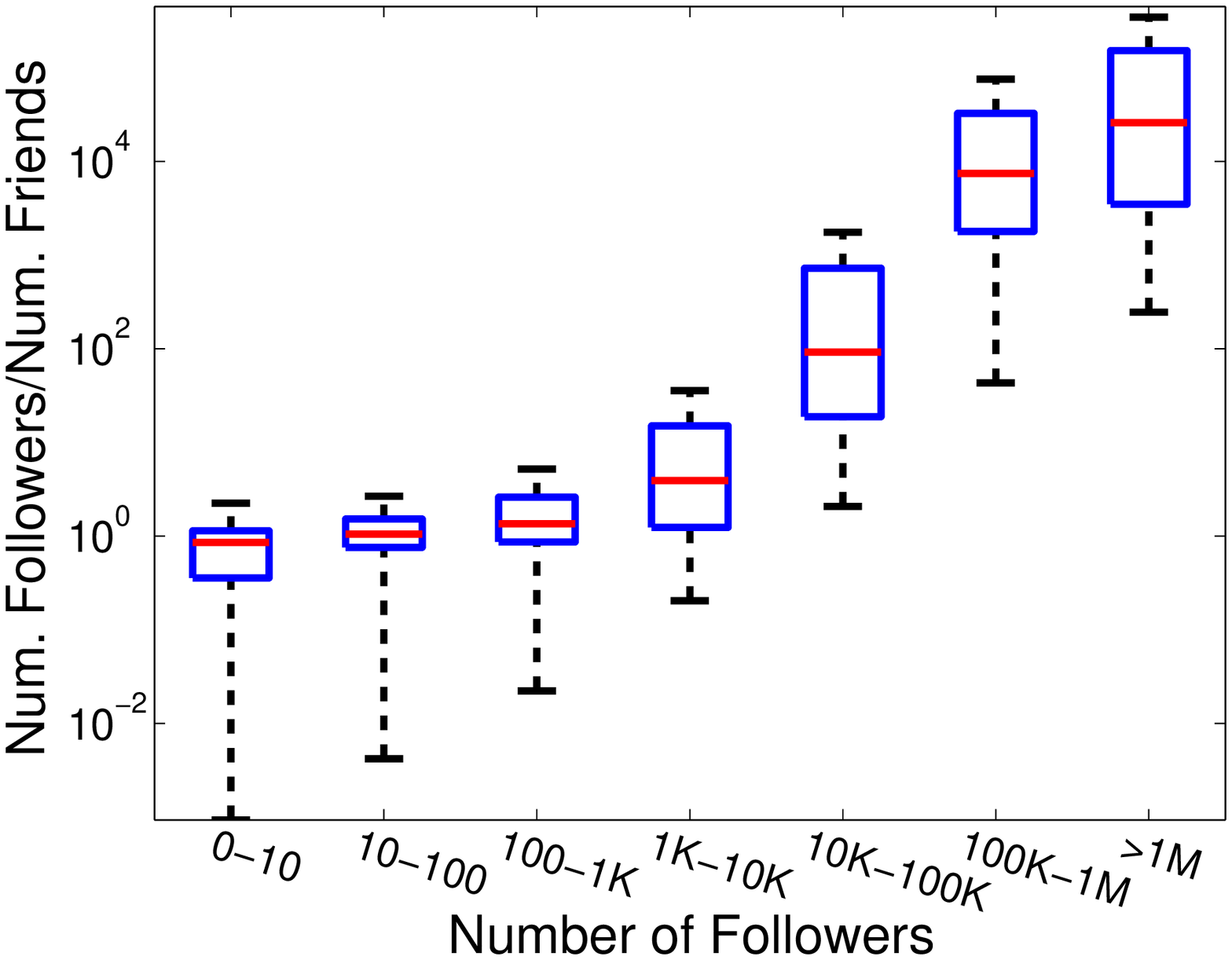}\label{fig:boxplot_frfo}}\hfill
	\subfigure[\% bidirectional relationships]{\includegraphics[width=0.5\textwidth]
	{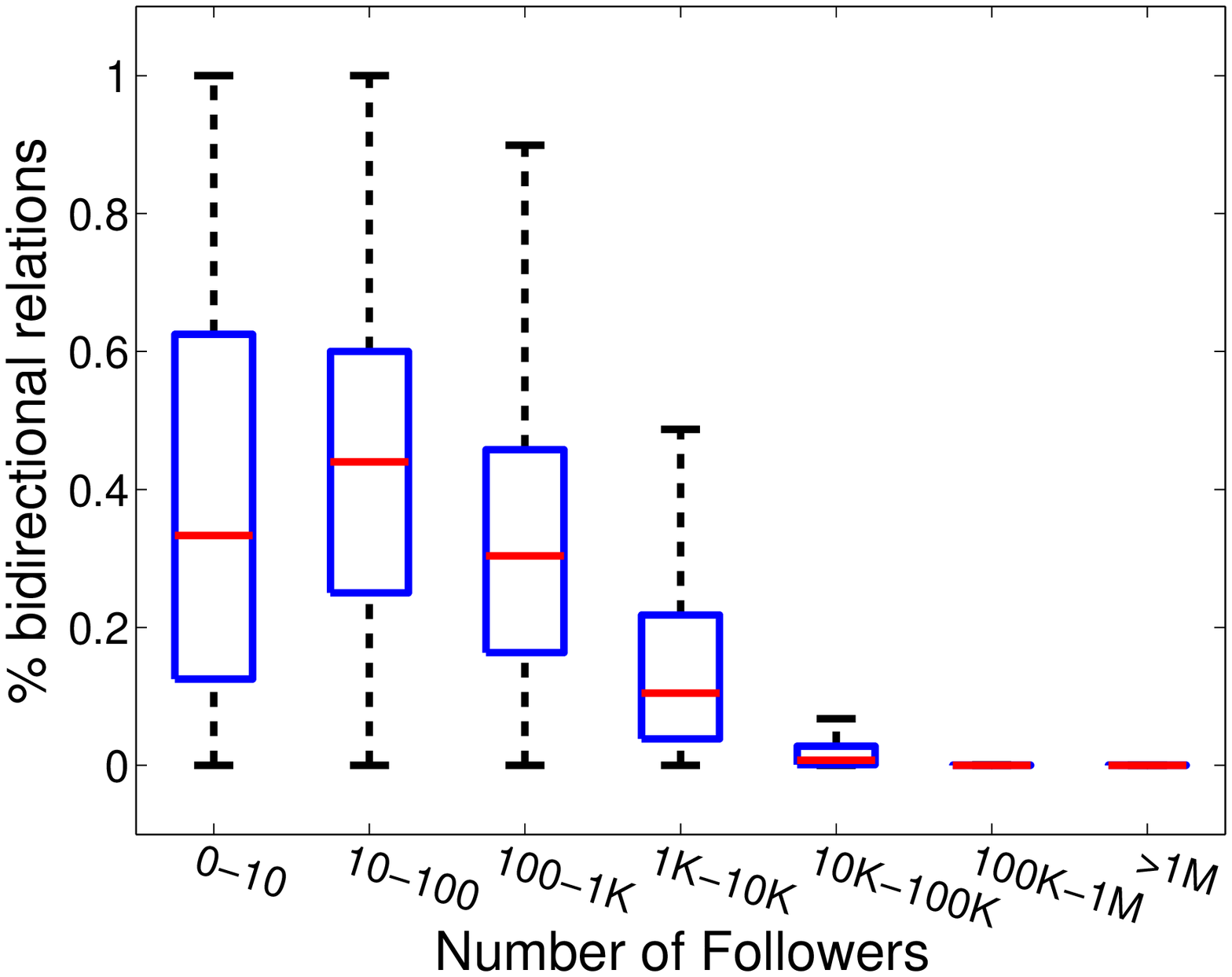}\label{fig:boxplot_bi}}
	\caption{The level of imbalance and reciprocation for different group of users based on their number of followers.}
	\label{fig:balance_vs_bidirectionality}
\end{minipage}
\vspace{-0.5cm}
\end{figure*}

\noindent
{\bf Public User Attributes:} 
We compare the willingness of users in different OSNs to publicly share their attributes in their profile. While this is not related to user activity, it is an indicator of user engagement and interest in an OSN.
Roughly 48\% of all the LCC users in G+ were providing at least one extra attribute in addition to their sex in April 2012. This ratio decreased and then stabilized around 44\% in our last few LCC snapshots. We further examine the distribution of the number of visible attributes across LCC users for different LCC snapshots and compare them with 480K random Facebook users (in FB-Pro dataset from Table \ref{tab:other_datasets}) in Figure \ref{fig:cdf_public_attributes}. We recall that there are 21 different attributes in Facebook profile. Figure \ref{fig:cdf_public_attributes} shows that the distribution for all LCC snapshots is identical. Also G+ users publicly share a much smaller number of attributes compared to Facebook users. In particular, half of the users publicly share at least 6 attributes on Facebook while less than 10\% of G+ users share 6 attributes. Twitter profile only has 6 attributes and 3 of them are mandatory. The examination of TW-Pro dataset shows that 69\% and 13\% of Twitter users share 0 and 1 non-mandatory attribute, respectively. In short, G+ users appear to share more public and non-mandatory attributes than Twitter users but significantly less than Facebook users.

{\em In summary, the analysis of different aspects of user activity in G+ resulted in the following important points: 
{\em (i)} The number of active LCC users has steadily grown but roughly 60 times slower than the whole LCC population. {\em (ii)} Around 10\% of LCC users generate a majority of all posts and only 1/10th of these users receive most of the reactions of any type to their posts. This is due to the fact that the rate of received reactions is strongly correlated with the user posting rate. {\em (iii)} The comparison of user activity for G+ with Facebook and Twitter revealed that Facebook and Twitter users exhibit a higher rate of generating posts.}

%% file: connectivity-rr2.tex
\section{LCC Connectivity \& Its Evolution}
\label{sec:connect}
In this section, we focus on the evolution of different features of connectivity among LCC users over time as the system becomes more populated, and compare these features with other OSNs.


\noindent
{\bf Degree Distribution:}
The node degree distribution is one of the basic features of connectivity. Since G+ structure is a directed graph, we separately examine the distribution of the number of followers in Figure \ref{fig:cdf_followers} and friends in Figure \ref{fig:cdf_friends}. Each figure shows the corresponding distribution across users in each one of our LCC snapshots,  among Twitter users in TW-Con snapshot, and the distribution of neighbors for random Facebook users in FB-Con snapshots\footnote{Note that Facebook forces bidirectional relationships. Therefore, the distribution for Facebook in both figures is the same.}. This figure demonstrates a few important points:
First, the distributions of followers and friends for G+ users can be approximated by a power law distribution with $\alpha$ = 1.26 and 1.39 in LCC-Nov snapshot, respectively. A similar property has been reported for the degree distribution of other OSNs including Twitter \cite{Twitter_WWW2010}, RenRen \cite{Chinese_OSN}, and Flickr or Orkut \cite{Mislove_IMC07}. 
Second, comparing the shape of the distribution across different LCC snapshots, we observe that both distributions look very similar for all LCC snapshots. The only exception is the earliest LCC snapshot (LCC-Dec) that has a less populated tail. This comparison illustrates that the shape of both distributions has initially evolved as the LCC became significantly more populated and users with larger degree appeared, and then the shape of distributions has stabilized in recent months.  
Third, interestingly, the shape of the most recent distribution of followers and friends for G+ users is very similar to the corresponding distribution for Twitter users. The only difference appears in the tail of the distribution of number of friends which is due to the limit of 5K friends imposed by G+ \cite{max_num_users_g+}. 
{\em The stability of the distribution of friends and followers for G+ users in recent months coupled with their striking similarity with these features in Twitter indicates that the degree distribution for G+ network has reached a level of maturity.}
Fourth, while the distributions for Facebook are not directly comparable due to its bidirectional nature, Figure \ref{fig:degree_distribution} shows that the distribution of degree for Facebook users does not follow a power law \cite{Facebook1} as they generally exhibit a significantly larger degree than Twitter and G+ users. Specifically, 56\% of Facebook users have more than 100 neighbors while only 3.6\% (and 0.8\%) of the G+ (and Twitter) users maintain that number of friends and followers. 

\noindent
{\bf Balanced Connectivity \& Reciprocation:}
Our examination shows that the percentage of bidirectional relationships between LCC users has steadily dropped from 32\% (in Dec 2011) and became rather stable in recent months around 21.3\% (in Nov 2012). 
Again, we observe that this feature of connectivity among LCC users in G+ seems to have reached a quasi-stable status after the system have experienced a major growth. Interestingly, Kwak et al. \cite{Twitter_WWW2010} reported  a very similar fraction of bidirectional relationships (22\%) in their Twitter snapshot from July 2009. This reveals yet another feature of G+ connectivity that is very similar to the Twitter network and very different from the fully bidirectional Facebook network.
In order to gain a deeper insight on this aspect of connectivity, we examine the fraction of bidirectional connections for individual nodes and its relation with the level of (im)balance between node indegree and outdegree. This in turn provides a valuable clue about the user level connectivity and reveals whether users exchange or simply relay information. To quantify the level of balance in the connectivity of individual nodes, Figure \ref{fig:boxplot_frfo} plots the summary distribution of the ratio of followers to friends (using boxplots) for different group of users based on their number of followers in our most recent snapshot (LCC-Nov).
This figure demonstrates that only low degree nodes (with less than 100 followers) exhibit some balance between their number of followers and friends. Otherwise, the number of friends among G+ users grows much slower than the number of followers.

We calculate the percentage of bidirectional relationships for a node $u$, called $BR(u)$, as expressed in Equation \ref{eq:bidirectional} where Friend(u) and Follower(u) represent the set of friends and followers for u, respectively. In essence, $BR(u)$ is simply the ratio of the total number of bidirectional relationships over 
the total number of unique relationships for user $u$.
\begin{equation}
BR(u) = \frac{Friend(u) \cap Follower(u)}{Friend(u) \cup Follower (u)}
\label{eq:bidirectional}
\end{equation}
Figure \ref{fig:boxplot_bi} presents the summary distribution of $BR(u)$ for different groups of G+ users in LCC the based on their number of followers using the LCC-Nov snapshot. The results for other recent LCC snapshots are very similar. 
As expected, popular users ($>$ 10k followers) have a very small percentage of bidirectional relationships. As the number of followers decreases, the fraction of bidirectional relationships slowly increases until it reaches around 40\% for low-degree users ($<$ 1K followers). In short, even low degree users that maintain a balanced connectivity, do not reciprocate more than 40\% of their relationships.
Our inspection of 5\% of LCC users who reciprocate more than 90\% of their edges revealed that 
90\% of them maintain less than 3 friends/followers and less than 5\% of them have any public posts. These results collectively suggest that G+ users reciprocate a small fraction of their relationships which is often done by very low degree users with no activity. 

\begin{figure}[t]
\begin{minipage}{0.48\columnwidth}
	\includegraphics[width=\columnwidth]
	{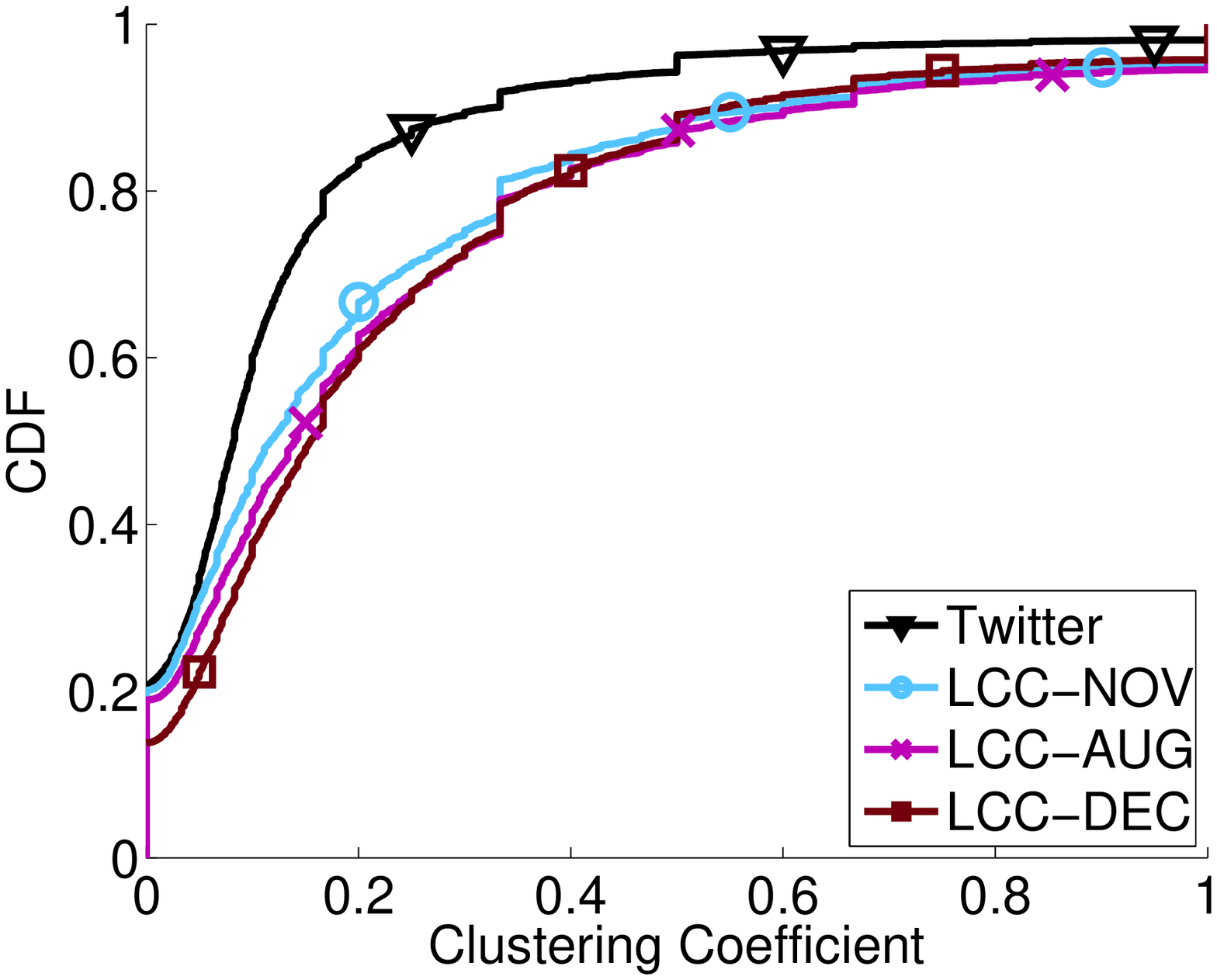}
	\caption{Clustering Coefficient}
	\label{fig:clust_coeff}
\end{minipage}	
\hfill
\begin{minipage}{0.48\columnwidth}
	\centering
	\includegraphics[width=\columnwidth]
	{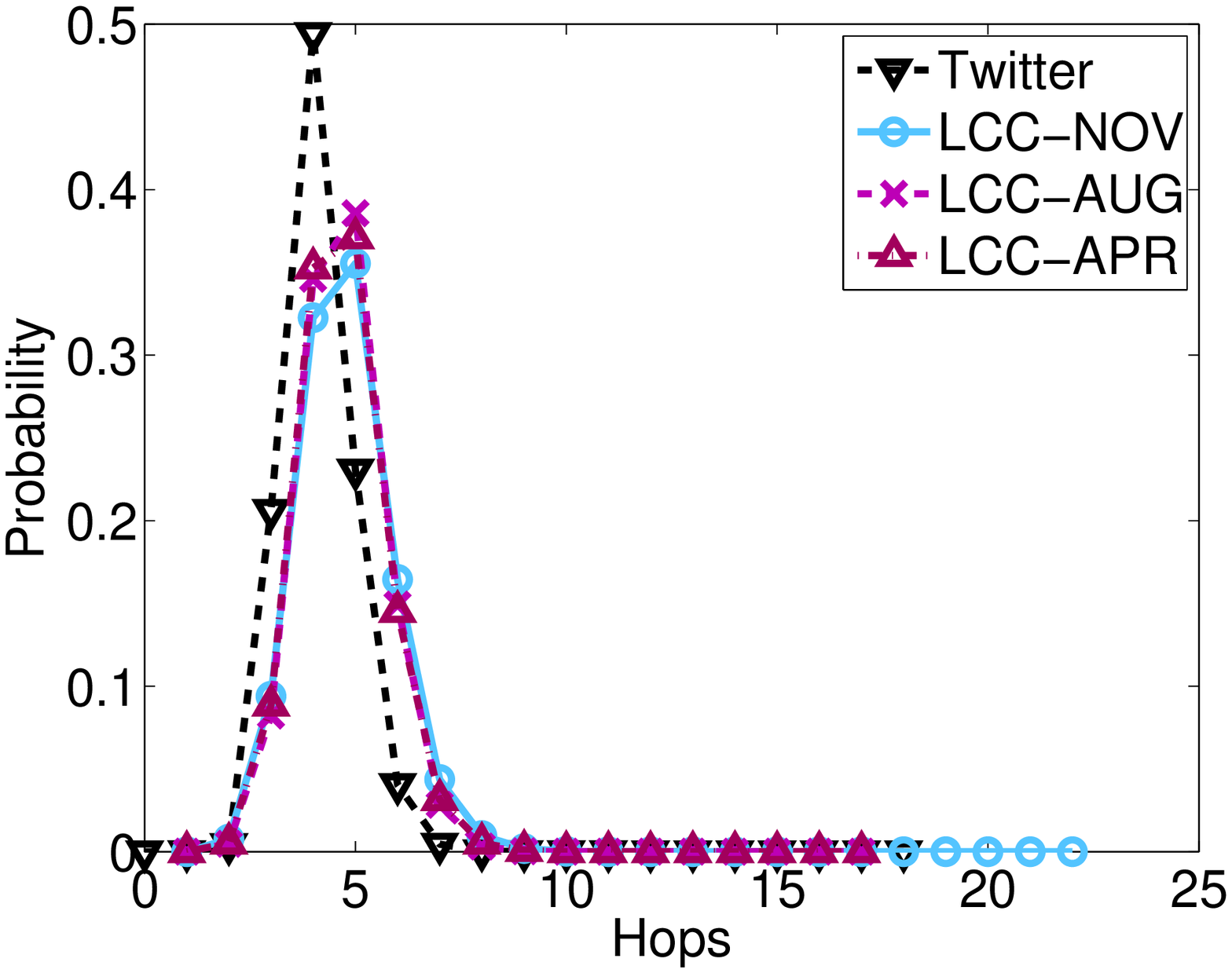}
	\caption{Average Path Length}
	\label{fig:avg_path_length}
\end{minipage}
\vspace{-0.2cm}
\end{figure}

\noindent
{\bf Clustering Coefficient:} Figure \ref{fig:clust_coeff} depicts the summary distribution of the undirected version of the clustering coefficient (CC) among G+ users in different LCC snapshots.
This figure clearly illustrates that during the roughly one year period from Dec 2011 to Nov 2012, the CC among the bottom 90\% of users remained below 0.5 and continuously decreases. On the other hand, the CC for the top 10\% of users has been very stable. In essence, {\em the G+ structure has become less clustered as new users joined the LCC over the one year period}. A similar trend in cluster coefficient has been recently reported for a popular Chinese OSN \cite{zhao2012multi} that indicates that such an evolution in the CC might be driven by underlying social forces rather than features of the OSNs. We also notice that the distribution of the CC among G+ users exhibits only minor changes between Aug and Nov 2012 which is another sign of stability in the connectivity features of G+ network.
Compared to Twitter network where the CC is less than 0.3 for 90\% of users, G+ is still more clustered. Furthermore, using the approximation presented in \cite{g+_imc_cha}, we conclude that just 1\% of the nodes in a complete Facebook snapshot collected in May 2011 \cite{Facebook1} have a CC larger than 0.2 in comparison with the 16\% and 30\% in Twitter and G+ (using LCC-Nov snapshot). In summary, as the population of G+ has grown, its connectivity has become less clustered but it is still the most clustered network compared to Twitter and Facebook.


\begin{table}
\small
\centering
\begin{tabular}{|c||c|c|c|c|c|c|}\hline
    & LCC-Nov & FB & Twitter\\ \hline\hline
Path Length (Avg) & 4.7 & 4.7 & 4.1\\
Path Length (Mode) & 5 & 5 & 4\\
Eff. Diameter & 6 & - &  4.8 \\
Diameter & 22 & 41 & 18 \\\hline
\end{tabular}
\caption{Summary of path length and diameter characteristics for G+, Facebook and Twitter}
\label{tab:apl}
\vspace{-0.5cm}
\end{table}

\noindent
{\bf Path Length:} Figure \ref{fig:avg_path_length} plots the probability distribution function for the pairwise path length between nodes in different LCC snapshots for G+ and a snapshot of Twitter (TW-Con). 
We observe that roughly 97-99\%  of the pairwise paths between G+ users are between 2 to 7 hops long and roughly 68-74\% of them are 4 or 5 hops. The diameter of the G+ graph has increased from 17 hops (in April) to 22 hops (in November of 2012). The two visibly detectable changes in this feature of the G+ graph as a result of its growth are: a small decrease in typical path length (from April to November) and the increase of its diameter in the same period.
Table \ref{tab:apl} summarizes the average and mode path length, the diameter and the efficient diameter \cite{leskovec2005graphs} (\ie,~90 percentile of pairwise path lengths) for the G+ network (using LCC-Nov), Twitter (using TW-Con) and a Facebook  snapshot from \cite{Facebook2}. We observe that G+ and Facebook exhibit similar average (and mode) path length but Facebook has a longer diameter. One explanation is the fact that the size of Facebook network is roughly one order of magnitude larger than G+ LCC. Twitter has the shortest average and mode path length and diameter among the three. We conjecture that this difference is due to the lack of restriction in the maximum number of friends that leads to many shortcuts in the network as Twitter users connect to a larger number of friends.

\noindent {\bf Relating User Activity \& Connectivity:}
We also analyzed the correlation between the connectivity and activity of individual users in the LCC. Our results reveal a strong positive correlation between the popularity of a user (\ie,~number of followers) and the user's post rate. The post rate of individual users exhibit a weaker correlation with the number of friends. Further details on all of our analysis can be found in our related technical report \cite{g+_rcuevas}.


%

{\em In summary, our analysis on the evolution of LCC connectivity led to the following key findings:
(i) As the size of LCC significantly increased over the past year, all connectivity features of LCC have initially evolved but have become rather stable in recent months despite its continued growth. (ii) Only low degree and non-active users may reciprocate a moderate fraction of their relationships. (iii) Many key features of connectivity for G+ network (\eg,~degree distribution, fraction of bidirectional relationships) have striking similarity with the Twitter network and very different from the Facebook network. These connectivity features collectively suggest that G+ is primarily used for message propagation similar to Twitter rather than pairwise users interactions similar to Facebook.}

%% file: related_work.tex
\section{Related Work}
\label{sec:rw}

\noindent \textbf{OSN characterization:} The importance of OSNs has motivated researchers to characterize different aspects of the most popular OSNs. The  graph properties of Facebook \cite{Facebook1,Facebook2}, Twitter \cite{Twitter_WWW2010,cha2010measuring} and other popular OSNs  \cite{Mislove_IMC07} have been carefully analyzed. Note that all these studies use a single snapshot of the system to conduct their analysis, instead we analyze the evolution of the G+ graph over a period of one year.  In addition, some other works leverage passive (e.g., click streams) \cite{benevenuto2009characterizing,schneider2009understanding} or active \cite{xu2012modeling,gyarmati2010measuring} measurements to analyze the user activity in different popular OSNs. These papers are of different nature than ours since they use smaller datasets to analyze the behaviour of individual users. Instead, we use a much larger dataset to analyze  the evolution of the aggregate public activity along time as well as the skewness of the contribution overall activity across users in G+.
Finally, few works have also analyzed users' information sharing through their public attributes in OSNs such as Facebook \cite{mislove2010you}.

\noindent \textbf{Evolution of OSN properties:} Previous studies have separately studied the evolution of the relative size of the network elements for specific OSNs (Flickr and Yahoo 360) \cite{kumarKDD}, the growth of an OSN and the evolution of its graph properties \cite{mislove2008growth,ahn2007analysis,zhao2012multi,gaito2012bursty,garg2009evolution,Reza_IEEENetwork} or the evolution of the interactions between users \cite{Chinese_OSN} and  users' availability  \cite{boutet2012impact}. In this paper, instead of looking at a specific aspect, we perform a comprehensive analysis to study the evolution of different key aspects of G+ namely, the system growth, the representative of the different network elements, the LCC connectivity and activity properties and the level of information sharing.

\noindent \textbf{Google+ Characterization}:
G+ has recently attracted the attention of the research community.
Mango et al. \cite{g+_imc_cha} use a BFS-based crawler to retrieve a snapshot of the G+ LCC between Nov and Dec 2011. They analyze the graph properties, the public information shared by users and the geographical characteristics and geolocation patterns of G+. Schiberg et al. \cite{schioberg2012tracing} leverage Google's site-maps to gather G+ user IDs and then crawl these users' information. In particular, they study the growth of the system and users connectivity over a period of one and a half months between Sep and Oct 2011. Unfortunately, as acknowledged by the authors the described technique was anymore available after Oct 2011. Furthermore, the authors also analyze the level of public information sharing and the geographical properties of users and links in the system. Finally, Gong et al. \cite{gong2012evolution} use a BFS-based crawler to obtain several snapshots of the G+ LCC in its first 100 days of existence. Using this dataset the authors study the evolution of the main graph properites of G+ LCC in its early stage. Our work presents a broader focus than these previous works since in addition to the graph topology and the information sharing we also analyze (for first time) the evolution of both the public activity and the representativeness of the different network elements. Furthermore, our study of the graph topology evolution considers a 1 year window between Dec 2011 and Nov 2012 when the network is significantly larger and presents important differences to its early status that is the focus of the previous works. In another interesting, but less related work, Kairam et al. \cite{kairam2012talking} use the complete information for more than 60K G+ users (provided by G+ administrators) and a survey including answers from 300 users to understand the selective sharing in G+. Their results show that public activity represents 1/3 of the G+ activity and that an important fraction of users make public posts frequently. Finally, other papers have studied the video telephony system of G+  \cite{xuvideo}, the public circles feature \cite{fang2012look} and the collaborative privacy management approaches \cite{hu2012enabling}.






\ 

%% file: conclusions.tex
\vspace{-0.7cm}
\section{Conclusion}
\label{sec:conclusion}

This paper examines the key features of G+ network and their evolution
during the first year of G+ operation. We conduct large scale measurement 
on G+ and collect some of the largest public datasets on any OSN to date
to characterize connectivity, activity and information sharing across G+ 
users along with their evolution over a one year period. We develop an 
efficient technique to collect random samples of G+ 
user. This in turn enables us to determine the relative size of key 
components (\ie LCC, partitions, singletons) of G+ network.

We show that while the size of LCC component of G+ has grown at a high 
rate (200K user per day), the relative size of LCC has decreased with 
time. Our investigations reveal that a significant fraction of new G+ 
users appear to be implicitly added by Google while they register for 
other Google services.
Furthermore, the main connectivity features of LCC have become relatively 
stable in recent months which suggests that the G+ network has 
reached a steady state.
We show that these stable connectivity features of LCC component of G+ 
have a striking similarity with Twitter but are very different from Facebook.
This similarity indicates that users use G+ for message propagation 
similar to Twitter rather than pairwise user interaction like Facebook.
In terms of user activity, even LCC users are not actively engaged in G+ network.
The contribution of user activity in terms of posting is skewed among LCC
users (\ie 10\% of users are responsible for 80\% of posts) and
user reactions to activities is an order of magnitude more skewed
(\ie 1\% of users generate 80\% of reactions to all posts).
Our findings collectively demonstrate that in the current OSN marketplace 
with two dominant players, namely Facebook and Twitter, a new OSN 
such as G+ might be able to attract a rather significant number of users 
to become part of the network (\ie connect to its LCC). However, it is
much more challenging to get these users meaningfully engaged in the 
system.

%% file: acknowledgments.tex
\section{Acknowledgements}

The authors would like to thank anonymous reviewers for their valuable feedback as well as Meeyoung Cha, Sue Moon, Diego Saez and Haewoon Kwak for sharing information from their datasets with us.
This work has been partially supported by the European Union through the FP7 TREND  (257740) and eCOUSIN (318398) Projects and the TWIRL (ITEA2-Call 5-10029) Project, the Spanish Goverment throught the MINECO eeCONTENT Project (TEC2011-29688-C02-02) and the MECD Jose Castillejo Grant (JC2011-0353), the Regional Government of Madrid through the MEDIANET project (S-2009/TIC-1468), the  Social Networks Chair of Institut-Mines Telecom SudParis  and the National Science Foundation under Grant IIS-0917381.